\documentclass[12pt]{article}
\usepackage{amsmath}
\usepackage{graphicx}
\usepackage{enumerate}
\usepackage{natbib}
\usepackage{url} 
\usepackage{makecell} 
\usepackage{amssymb}
\usepackage{multicol}
\usepackage{multirow}
\usepackage{xcolor}
\usepackage{amsthm}
\usepackage{setspace,comment}
\usepackage{hyperref,enumitem}
\usepackage{array,bm}
\usepackage{amsthm}
\theoremstyle{remark}

\theoremstyle{definition}
\newtheorem{definition}{Definition}
\newcommand{\blind}{1}
\newcommand{\dd}{\text{d}}

\newcommand{\ff}{\mathsf f}

\begin{document}

\def\spacingset#1{\renewcommand{\baselinestretch}%
{#1}\small\normalsize} \spacingset{1}


\if1\blind
{
  \title{\bf Second-order spatial analysis of shapes of tumor cell nuclei}
  \author{Ye Jin Choi, Sebastian Kurtek\\
    Department of Statistics, The Ohio State University\\
    Simeng Zhu\\
    James Cancer Center, The Ohio State University\\
    Karthik Bharath \\
    School of Mathematical Sciences, University of Nottingham}
  \maketitle
} \fi

\if0\blind
{
  \bigskip
  \bigskip
  \bigskip
  \begin{center}
    {\LARGE\bf Second-order spatial analysis of shapes of tumor cell nuclei}
\end{center}
  \medskip
} \fi

\bigskip
\begin{abstract}
Intra-tumor heterogeneity driving disease progression is characterized by distinct growth and spatial proliferation patterns of cells and their nuclei within tumor and non-tumor tissues. A widely accepted hypothesis is that these spatial patterns are correlated with morphology of the cells and their nuclei. Nevertheless, tools to quantify the correlation, with uncertainty, are scarce, and the state-of-the-art is based on low-dimensional numerical summaries of the shapes that are inadequate to fully encode shape information. To this end, we propose a marked point process framework to assess spatial correlation among shapes of planar closed curves, which represent cell or nuclei outlines. With shapes of curves as marks, the framework is based on a mark-weighted $K$ function, a second-order spatial statistic that accounts for the marks' variation by using test functions that capture only the shapes of cells and their nuclei. We then develop local and global hypothesis tests  for spatial dependence between the marks using the $K$ function. The framework is brought to bear on the cell nuclei extracted from histopathology images of breast cancer, where we uncover distinct correlation patterns that are consistent with clinical expectations. 
\end{abstract}

\noindent%
\vfill

\newpage
\spacingset{1.55} 

\section{Introduction}
\label{sec:intro}

The tumor microenvironment (TME) comprises tumor cells, stromal (supportive) cells, and other components, together driving abnormal, rapidly growing structures \citep{hanahan2012accessories}. While both the tumor cell and its nucleus offer valuable information for tumor characterization, research has focused on studying the nucleus due to its more direct association with genetic alterations \citep{hanahan2011hallmarks}. Further, extracting cell boundaries from histopathology images is often unreliable, owing to overlapping cells and poorly defined cytoplasmic edges. Tumor cell nuclei, unlike those of normal cells, frequently exhibit morphological abnormalities such as indentations, folds and fragmentation across various cancer types \citep{singh2022nuclear}. These nuclear irregularities are not only visually distinctive, but also serve as strong predictors of cancer progression and treatment outcomes \citep{nafe2005morphology,de2011large}.

Pathologists have long relied on nuclear morphology for diagnostic purposes. In the specific context of breast cancer, histopathological examination of cancer tissue provides critical prognostic information that guides clinical treatment decisions. Histologic grading serves as a standardized assessment tool whereby pathologists evaluate tumor specimens from biopsies or surgical excisions. The current Nottingham grading system incorporates three key morphological parameters: (i) the extent of tubule formation, (ii) the degree of nuclear pleomorphism, and (iii) mitotic activity count \citep{Bloom1957,1991_Histopathology_vol19_no5}. However, the inherent subjectivity in manually assessing these morphological features has raised concerns regarding inter-observer reproducibility, prompting efforts to develop more objective, quantitative assessment methods \citep{van2022grading}.

As a result, many recent studies have relied on quantitative approaches for evaluating morphological variation in tumor cell nuclei, derived from histopathology images, to pursue several scientific and clinical goals, including automatic differentiation of malignant and benign cells \citep{fischer2020nuclear}, classification of tumor subtypes \citep{beck2011systematic}, prediction of cancer prognosis and patient survival \citep{murphy1990nuclear}, discovery of associations with genetic mutations \citep{sali2024morphological}, and forecasting of patient response to chemotherapy \citep{kather2019predicting}. Further, numerous studies have indicated that spatial interactions among cells in the TME also help differentiate cancer grades \citep{barua2018functional,tsujikawa2020prognostic,wang2022spatial}. However, these studies focused on cell locations only. In contrast, \cite{sali2024morphological} and \cite{lu2020prognostic} emphasized that intra-tumor or local morphological diversity of tumor cells' nuclei is a key cancer trait and a strong prognostic indicator. To capture their shape, \cite{singh2010analysis} represented the three-dimensional structure of stromal cells using spherical harmonics and explored principal modes of shape variation in relation to duct proximity. However, they only distinguished between `close' or `far' cells. Similarly, \cite{nafe2005histomorphometry} used various shape descriptors and size, along with spatial information, e.g., neighboring nuclei or inter-nuclear distances, for tumor grade classification. In both cases, representing the shape of cells or their nuclei using a finite number of spherical harmonics or subjectively chosen descriptors cannot fully capture their morphological complexity, resulting in potentially significant loss of information.

\subsection{Motivation and contributions}
This paper is motivated by the following questions.
\begin{enumerate}
    \item [(i)] Are the shapes of nuclei of tumor cells correlated with their spatial configurations within the tumor microenvironment?
    
    \item[(ii)] How does the morphological variation depend on the spatial proximity between normal or tumor cell nuclei?
\end{enumerate}

Answers to the two questions along with a procedure to quantify the accompanying uncertainty offer crucial empirical validation to clinical observations. For example, in the context of breast cancer to be considered in Section \ref{sec:real-data}, it is observed that normal breast cells arrange in a regular spatial tubular pattern around ducts while tumor cells form clusters based on similar genetic information \cite{GreavesMel2012Ceic}; our findings using the developed framework corroborate these assertions.

\begin{figure}[!t]
    \centering
\includegraphics[width=1\linewidth]{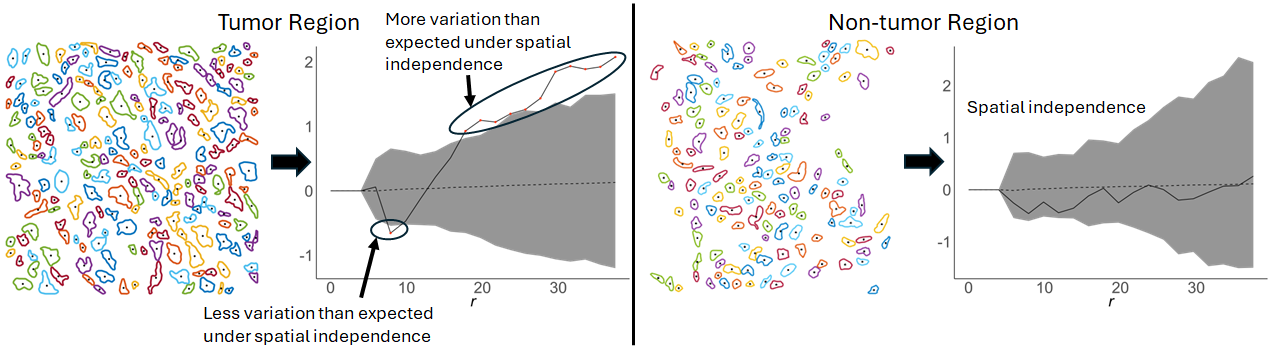}
    \caption{\small We contrast spatial correlation patterns at different scales among cell nuclei shapes within tumor (left) and non-tumor (right) regions via a second-order spatial statistic (solid black line). The shaded region is an envelope that captures underlying uncertainty.}
    \label{fig:mot_fig}
\end{figure}

Motivated by and tailored to tackle the two questions arising in the quantitative study of intra-tumor heterogeneity in two-dimensional pathology images, we propose a spatial point process model for the cell nuclei with marks that encode shape-related characteristics of their outlines. 
This is sensible since the location and number of cells (and their nuclei) in a histopathology image are not known \emph{a priori}. The shape of a cell nucleus under the parameterized curve representation is a quantity unaffected by the nucleus’ location, size, orientation and parameterization, referred to as its symmetries. While the physical location and parameterization of the nuclei constitute nuisance variation, size and orientation may provide useful information to characterize their morphological diversity, e.g., as related to tumor growth processes. We provide an illustration of the framework in Figure \ref{fig:mot_fig}. We start with cell nuclei outlines within a region of interest. We then compute a second-order spatial statistic (black solid line), as described in Sections \ref{sec:Kfunction} and \ref{sec:hypothesis-testing}, that accounts for the cell nuclei' shape variation at different scales (defined through balls with different radii). In addition, we quantify uncertainty in the computed statistic by constructing an envelope (gray shaded region) via an appropriate bootstrap procedure. In the left panel, corresponding to a region selected inside the tumor, it is evident that the statistic deviates below the envelope at small scales indicating spatial correlation among the shapes of nuclei that manifests itself through less variation than expected under independence. At larger scales, the statistic deviates above the envelope indicating that the nuclei shapes are spatially correlated with more variation than expected if they were independent. In the right panel, which considers a region within non-tumor tissue, the nuclei shapes are uncorrelated in space since the statistic never deviates outside of the envelope. Thus, spatial correlation patterns for cell nuclei shapes within tumor and non-tumor tissues are distinct.

The correlation between coordinates of the point process and its marks may be studied using variants of Ripley’s $K$ function  developed for homogeneous \citep{ripley1976second} and inhomogeneous \cite{baddeley2000non} point processes. In the presence of marks, weights associated with each pair of spatial locations are estimated using appropriate \emph{test functions} \citep{illian2008statistical}, which include the mark correlation function \citep{stoyan1994fractals} (summarizing average numerical similarity), the mark variogram \citep{cressie2015statistics} (summarizing average variability as a function of spatial distance), and Moran’s I \citep{moran1950notes} (summarizing covariance); see \cite{illian2008statistical} for a more comprehensive discussion. Depending on the chosen test function, one obtains different second-order summary measures. A variety of marks has been studied in the statistics literature, including integer-valued (qualitative, multi-type or multivariate) \citep{lotwick1982methods, van1999indices}, real-valued \citep{schlather2001second, schlather2004detecting, lieshout2006aj}, or mixed integer- and real-valued \citep{eckardt2019analysing}. 

Extensions to function-valued marks appear in \cite{comas2008analysing, comas2011second, comas2013analysing}. Most recently, \cite{ghorbani2021functional} introduced a comprehensive framework that incorporates functional data with auxiliary variables as marks, proposing a mark-weighted $n$th-order reduced moment measure (based on test functions) to generalize existing summary statistics. In particular, they considered a mark-weighted inhomogeneous $K$ function to quantify spatial correlatedness of the marks, while adjusting for the varying intensity of the spatial point process. Other extensions consider multivariate function-valued marks \cite{eckardt2024marked} and compositional marks \cite{eckardt2025spatial}. 

The shape setting considered in this paper lies outside the capabilities of the above works with functional marks. Accordingly, we
\begin{enumerate}
\itemsep 0em
    \item [(a)] define a novel population mark-weighted $K$ function that accounts for the shapes of cell nuclei outlines as functional marks with symmetries, and a corresponding estimator that is easy to compute and is unbiased under some conditions; 
    \item [(b)] use the estimator to construct global and local envelope tests for spatial dependence of the marks, which help answer questions (i) and  (ii) above. 
\end{enumerate}
In the present setting, any test function used to define a mark-weighted $K$ function must be invariant to the symmetries of translation, rescaling, rotation and reparameterization. The key issue lies in ensuring invariance to reparameterizations, which constitute an infinite-dimensional Lie group. This is addressed by using an elastic Riemannian metric, invariant to reparameterizations, that then results in a nonlinear functional shape (or size-and-shape or orientation-and-shape) mark space. At the population level, the mark-weighted $K$ function is defined using a product measure on the space of closed curves and desired symmetries, which at the sample level leads to a simple estimator. 

The rest of this manuscript is organized as follows. Section \ref{sec:elastic-shape-analysis} reviews elastic shape analysis of closed curves. Section \ref{sec:fmpp} defines a marked spatial point process wherein marks are closed curves with desired symmetries. Section \ref{sec:Kfunction} introduces the shape, size-and-shape and orientation-and-shape mark-weighted $K$ functions and provides a recipe for their estimation. Section \ref{sec:hypothesis-testing} defines a formal hypothesis test of independence between marks and their spatial locations wherein the proposed mark-weighted $K$ functions serve as test statistics. Section \ref{sec:simul} presents simulations that assess the proposed framework. In Section \ref{sec:real-data}, the framework is applied to cell nuclei outlines derived from histopathology images of breast cancer. Finally, we close with a brief discussion in Section \ref{sec:discussion}. The supplement contains (i) a discussion of the reference measure on the mark space of shapes (Section 1), (ii) a derivation of the normalizing constant for the proposed mark-weighted $K$ function (Section 2), (iii) a proof that the proposed $K$ function estimator is unbiased (Section 3), (iv) a brief discussion of challenges associated with establishing consistency of the $K$ function estimator (Section 4), (v) four additional histopathology images used in our analysis (Section 5), (vi) an additional result for cell nuclei from a region sampled in non-tumor tissue (Section 6), and (vii) code and data used to compute the results presented in Sections \ref{sec:simul} and \ref{sec:real-data}.

\section{Comparing curves using elastic metrics}
\label{sec:elastic-shape-analysis}

In this section, we briefly review the elastic framework for comparing planar parameterized curves, and refer to \cite{srivastava2016functional} for details.

\subsection{Curve representation and symmetries}

Key to quantifying correlations between shapes of cell nuclei and their locations within the tissue is to systematically compare nuclei contours. This requires establishing correspondence between points on two or more distinct curves. To this end, we represent cell nuclei as closed parameterized planar curves belonging to the set $\mathcal B:=\{\beta: \mathbb S^1 \to \mathbb R^2 \mid \beta \text{ is absolutely continuous}\}$. The benefits of comparing objects given by planar outlines under a parameterized curve representation instead of just a subset of $\mathbb R^2$ are now well-understood \citep[Chapters 6-7][]{srivastava2016functional}. 

As discussed earlier, the class of relevant symmetries associated with cell nuclei outlines will depend on many factors, including the TME and its location within the tissue. There are four transformations of a parameterized curve $\beta$ that constitute symmetries pertaining to its shape: translation, scaling, rotation and reparameterization. Given a curve $\beta \in \mathcal B$, for every $t \in \mathbb S^1$, its translation is $\beta(t)+a$ where $a \in \mathbb R^2$, its scaling is $\sigma \beta(t)$ with $\sigma>0$, its rotation is $O\beta(t)$, where $O \in SO(2) = \{O \in \mathbb{R}^{2\times 2} \mid OO^\top = O^\top O = I_2, \text{det}(O) = 1\}$; its reparameterization is the function composition $\beta \circ \gamma$, where $\gamma:\mathbb S^1 \to \mathbb S^1$ is a diffeomorphism of $\mathbb S^1$, such that the image of $\beta \circ \gamma$ equals that of $\beta$ for every such  $\gamma$. Since reparameterizing $\beta$ representing the outline of a cell nucleus preserves its image, it ought to be considered as a symmetry for curves, \emph{irrespective} of any of the other factors that may influence the spatial configuration of the cell nuclei.  In fact, the set
$\Gamma:=\{\gamma:\mathbb S^1 \to \mathbb S^1 \mid \gamma \text{ is a diffeomorphism}\}$
is an infinite-dimensional Lie group, with composition as the group operation. The shape symmetries then are represented by the product group, referred to as the \emph{symmetry group}, $\mathbb R^2 \rtimes (\mathbb R_+ \times SO(2) \times \Gamma)$, with the semi-direct product action 
$((a,\sigma,O,\gamma),\beta) \mapsto \sigma O \beta \circ \gamma +a$. The order in which one scales, rotates and reparameterizes a curve is immaterial, since the individual symmetries commute. After these operations, a translation is applied.

The sizes of cell nuclei plausibly influence their spatial scatter, and it may be desirable then to not treat scale as a symmetry. In this case, the symmetry group is $\mathbb R^2 \rtimes (SO(2) \times \Gamma)$. Alternatively, the orientation of the nuclei might encode relevant information on infiltration of the tumor cell in a region of the tissue, and thus not nuisance; in this case, the symmetry group is $\mathbb R^2 \rtimes (\mathbb R_+  \times \Gamma)$. As such, we will consider three symmetry groups: (i) translation, scale, rotation and reparameterization, (ii) translation, rotation and reparametrization, and (iii) translation, scale and reparameterization. We generically denote a symmetry group by $G$, and refer to its elements as \emph{shape symmetries}, since the image-preserving reparametrization group is always considered to be a symmetry. We denote the action of a symmetry $g$ on a curve $\beta$ by $\beta*g$. Upon choosing $G$, the next step is to compare curves in a manner that is invariant to $G$. In other words, given two curves $\beta_1$ and $\beta_2$, a distance $d$ such that $d(\beta_1 * g, \beta_2 *g)=d(\beta_1,\beta_2)$ for every $g \in G$ is desired. 

\subsection{Elastic metric and the square-root velocity transform}

The geometric framework \citep{joshi2007novel,mio2007shape} based on the \emph{elastic Riemannian metric} on $\mathcal B$ provides a convenient set of tools to define a distance on curves compatible with the required symmetries. Definition of the metric requires a fair amount of preparatory notation, and thus, for ease of exposition, we refer the reader to Chapter 5 of \cite{srivastava2016functional}. Summarily, the metric quantifies differences between two curves $\beta_1$ and $\beta_2$ by measuring the minimal deformation needed for bending, with weight $a>0$, and stretching, with weight $b>0$, whilst being invariant to reparameterization and rotation. This is a key requirement, since the parameterization of a curve, regardless of the chosen symmetry group, is nuisance. However, computing the distance using the elastic Riemannian metric for any choice of weights $a,b$ is difficult. The situation is salvaged by the Square-Root Velocity (SRV) transform of a curve $\beta \in \mathcal B$, defined as
\begin{eqnarray}
	q(t) :=\begin{cases} 
		\frac{\dot{\beta}(t)}{\sqrt{|\dot{\beta}(t)|}} & \text{if } |\dot{\beta}(t)| \neq 0, \\
		0 & \text{otherwise,}
	\end{cases}
\end{eqnarray}
where $\dot \beta(t)=\frac{\dd \beta(t)}{\dd t}$ is the velocity vector. When $\beta \in \mathcal B$, its SRV transform $q$ satisfies $\|q\|^2:= \int_{\mathbb S^1} |q(t)|^2 \dd t <\infty$, where $|\cdot|$ is the standard norm corresponding to the inner product $\langle\cdot,\cdot\rangle$ in $\mathbb R^2$, and $\|\cdot\|$ is the norm on $\mathbb L^2(\mathbb S^1, \mathbb R^2)$ arising from the inner product $\langle\langle\cdot,\cdot\rangle\rangle$. Moreover, since $\beta$ is a closed curve, a constraint imposed on  $q$ is $\int_{\mathbb{S}^1} \dot{\beta}(t) \, \dd t = \int_{\mathbb{S}^1} q(t) |q(t)| \, \dd t = 0$. 

Let $\mathbb{L}^2(\mathbb S^1, \mathbb R^2)$ be the Hilbert space of square-integrable functions from the unit circle to the plane. The relevant function space is then $\mathcal Q:=\left\{q \in \mathbb{L}^2(\mathbb S^1, \mathbb R^2) \mid \int_{\mathbb{S}^1} q(t) |q(t)| \, \dd t = 0 \right\}$,
which is a co-dimension two nonlinear submanifold of the infinite-dimensional Hilbert space $\mathbb{L}^2(\mathbb S^1, \mathbb R^2)$ \citep[p.173][]{srivastava2016functional}. When using the weights $a=1/4$ and $b=1$ in the elastic metric on $\mathcal{B}$ the elastic metric reduces to the standard $\mathbb{L}^2$ metric on $\mathcal Q$ under the SRV transform so that the distance between $q_1$ and $q_2$ invariant to the chosen symmetry group can be defined by using the usual $\mathbb L^2$ distance \citep{joshi2007novel}. 

\subsection{Curve correspondence and distance under symmetries}

Location/translation information of $\beta \in \mathcal B$ is lost when using its SRV transform $q$. When scale of $q$ is a symmetry, then its normed version ${q}\|q\|^{-1}$ can be considered; rotation and reparameterization symmetries, however, need to be accounted for when comparing $\beta_1$ and $\beta_2$ via their SRV transforms. The action of a rotation $O$ on $\beta$ is identical to that on $q$. However, when $\beta$ is reparametrized as $\beta \circ \gamma$ for $\gamma \in \Gamma$, the corresponding action of $\gamma$ on its SRV transform $q$ is given by $(q,\gamma):=(q \circ \gamma) \sqrt{\dot \gamma}$. 

To establish correspondence between curves in $\mathcal B$ invariant to a symmetry group $G$, we define a distance on the quotient $\mathcal Q/G$ consisting of equivalence classes or orbits of $q$ under the action of $G$. Curves invariant to their symmetries together are represented as points in quotient spaces, defined as follows. 
\begin{itemize}
	\itemsep-0.5em
	\item \emph{Shape space} with $G=SO(2) \times \Gamma$:\\	
	$\mathcal Q_{\text{sh}}:=\{[q]_{\text{sh}}, q \in \mathcal Q\}$ with $[q]_{\text{sh}}=\left\{q * g:=O\left(q\|q\|^{-1},\gamma\right)|(O,\gamma) \in G\right\}$.
	\item \emph{Size-and-shape space} with $G=SO(2) \times \Gamma$:\\
	$\mathcal Q_{\text{sc-sh}}:=\{[q]_{\text{sc-sh}}, q \in \mathcal Q\}$, with $[q]_{\text{sc-sh}}=\left\{q * g:=(Oq,\gamma)|(O,\gamma) \in G\right\}$.
\item 	\emph{Orientation-and-shape space} with $G=\Gamma$:\\
$\mathcal Q_{\text{ro-sh}}:=\{[q]_{\text{ro-sh}}, q \in \mathcal Q\}$, with $[q]_{\text{ro-sh}}=\left\{q * g:=\left(q\|q\|^{-1},\gamma\right)|\gamma\in G \right\}$.
\end{itemize}
The shape of $q$ is defined as its equivalence class under the action of $SO(2)\times\Gamma$, such that $q\|q\|^{-1}$ and $O(q\|q\|^{-1},\gamma)$ have the same shape for any $(O,\gamma) \in SO(2)\times\Gamma$; size-and-shape and orientation-and-shape of $q$ have analogous interpretations. 

We now define distances on the three spaces, which will eventually enables us to quantify how cell nuclei and their symmetries influence their spatial configuration. Their definitions are based on the observation that, for every $(O, \gamma) \in SO(2) \times \Gamma$, 
$\|O(q,\gamma)\|^2=\int_{\mathbb S^1} \left \langle Oq(\gamma(t))\sqrt{\dot \gamma(t)},Oq(\gamma(t))\sqrt{\dot \gamma(t)} \right \rangle \dd t=\int_{\mathbb S^1} |q(u)|^2\dd u=\|q\|^2$, so that the rotation and parameterization groups act by isometries on $\mathcal Q$. This ensures that suitable distances may be defined on the shape, size-and-shape and orientation-and-shape spaces, as follows. The extrinsic shape distance between two curves $q_1, q_2 \in \mathcal Q$ is defined as $d_{\text{sh}}(q_1,q_2) = \underset{(O,\gamma)\in SO(2) \times\Gamma}{\inf}\left\|q_1\|q_1\|^{-1}-O\left(q_2\|q_2\|^{-1},\gamma\right)\right\|$, and is a distance on the shape space $\mathcal Q_{\text{sh}}$. In similar fashion, the size-and-shape distance is
$d_{\text{sc-sh}}(q_1,q_2)=\underset{(O,\gamma)\in SO(d)\times\Gamma}{\inf} \; \left\| q_1 - O \left(q_2, \gamma\right) \right\|$,
interpreted as a distance on the size-and-shape space $\mathcal Q_{\text{sc-sh}}$. Finally, the extrinsic orientation-and-shape distance is defined as
$d_{\text{ro-sh}}(q_1,q_2)=\underset{\gamma\in\Gamma}{\inf}  \left\|q_1\|q_1\|^{-1}-(q_2\|q_2\|^{-1},\gamma)\right\|$,
viewed as a distance on the orientation-and-shape space $\mathcal Q_{\text{ro-sh}}$. We refer to the three distances on curves generically as \emph{shape distances}. 

The distances $d_{\text{sh}}$ and $d_{\text{ro-sh}}$ are referred to as extrinsic since they are based on the chordal distance on the infinite-dimensional sphere within $\mathcal Q$. Strictly speaking, $d_{\text{sh}}, d_{\text{sc-sh}},$ and $d_{\text{ro-sh}}$ are bonafide distance functions only if each equivalence class is closed; this is guaranteed by taking the closure of each equivalence by enlarging the group $\Gamma$ to a monoid by including reparameterizations that are weakly increasing \citep[Section 5.5][]{srivastava2016functional}. Computing the distances requires optimizing over $SO(2)$ and $\Gamma$. For a fixed $\gamma$, the solution to optimizing over $SO(2)$ is available in closed form via Procrustes analysis.  For a fixed $O$, a dynamic programming algorithm, with an additional seed search over starting points on $\mathbb{S}^1$, is used to optimize over $\Gamma$ \citep{robinson2012functional}. Joint optimization over rotations and reparameterizations is carried out by alternating between the two. 

A consequence of computing the distances above between a curve $q_1$ and another curve $q_2$ is the determination of the optimal element $\hat g$ of the symmetry group $G$ that best `matches' $q_1$ to $q_2*\hat g$. For example, if the shape distance $d_{\text{sh}}(q_1,q_2) = \|q_1\|q_1\|^{-1}-\hat O\left(q_2\|q_2\|^{-1},\hat \gamma\right)\|$ is realized at $(\hat O,\hat \gamma)$, then, with respect to the shape distance, $q_1\|q_1\|^{-1}$ is optimally aligned, or in shape correspondence, with $\hat O\left(q_2\|q_2\|^{-1},\hat \gamma\right)$. 

\subsection{Computing means and geodesics}
\label{subsec:mean_geo}	
Higher-order moment measures for point processes with marks require both, pairwise and multiple comparison of marks. While the former is carried out for curves as marks by computing the shape distances above, a template is required to establish correspondence between multiple curves. A natural choice for a template is a Karcher mean. 

Using the shape distances $d_j,\ j \in \{\text{sh},\text{sc-sh},\text{ro-sh}\}$, define the Karcher mean $\mu_{j}$ of a sample $q_1,\ldots,q_n$ of curves in $\mathcal Q$ as the minimizer of 
$q \mapsto \underset{q \in \mathcal Q}{\text{argmin}} \sum_{i=1}^n d_{j}(q,q_i)^2$. By construction, if $\mu_j$ is the Karcher mean with respect to $d_j$ for $ j \in \{\text{sh},\text{sc-sh},\text{ro-sh}\}$, then so is $\mu_j * g$ for any $g$ in the corresponding symmetry group. For example, $O(\mu_{\text{sh}},\gamma)$ is a Karcher mean with respect to the shape distance $d_{\text{sh}}$ for any $(O,\gamma) \in SO(2) \times \Gamma$ if $\mu_{\text{sh}}$ is. The Karcher mean is thus an equivalence class. In practice, computing the Karcher mean and multiple alignment of sample curves to the Karcher mean are performed simultaneously through an iterative algorithm. This algorithm alternates between aligning the sample curves to the current estimate of the Karcher mean and updating the mean based on the newly aligned curves. Ultimately, this process yields the Karcher mean and the sample shapes aligned to it. For further details, see \cite{srivastava2016functional}.

Upon determining the optimal symmetry $\hat g$ when computing a distance $d_j$ between two curves $q_1$ and $q_2$, the geodesic between $q_1$ and $q_2$ is given by $(1-\tau)q_1+\tau (q_2*\hat g),\ 0 \leq \tau \leq 1$. For example, when using the size-and-shape distance $d_{\text{sc-sh}}$, suppose $d_{\text{sc-sh}}(q_1,O(q_2,\gamma))$ is minimized at $\hat g=(\hat O,\hat \gamma) \in SO(2) \times \Gamma$. The size-and-shape geodesic between $q_1$ and $q_2$ is given by $(1-\tau)q_1+\tau \hat O (q_2,\hat \gamma),\ 0 \leq \tau \leq 1$; by construction, the midpoint $0.5 q_1+0.5 \hat O (q_2,\hat \gamma)$ is then the Karcher mean $\mu_{\text{sc-sh}}$. 

\section{Point process with marks as curves with symmetries}
\label{sec:fmpp}
The goal is to quantify the correlation between cell nuclei modulo their symmetries (e.g., shape, size-and-shape) and their locations within a histopathology image. To this end, we model the image as a realization of a marked point process in a spatial domain, with marks assuming values in the space of curves compatible with a chosen set of symmetries.

\subsection{Spatial domain, mark space and reference measures}
Consider $\mathcal X \subseteq \mathbb R^2$ equipped with the standard Euclidean norm $|\cdot |$ with Borel sets $\mathbb B(\mathcal X)$.  Given a probability space $(\Omega,\mathcal F, \mathbb P)$, let $\Psi_{\text{gr}}=\{x_i\}_{i=1}^N,\ N \in \{0,1,\ldots,\infty\}$ on $(\mathcal X, \mathbb B(\mathcal X))$ be a point process, referred to as the \emph{ground process}. Denote by $G$ one of the symmetry groups: $SO(2) \times \Gamma$ for shape and size-and-shape, and $\Gamma$ for orientation-and-shape. 

We work with the space $\mathcal Q$ containing SRV transformed curves from $\mathcal B$, and its quotients. Upon choosing a symmetry group $G$ for the curves representing nuclei contours, there are two candidates for the mark space $\mathcal M$: (i) the product space $\mathcal Q \times G$, or (ii) a quotient of $\mathcal Q$, such as the shape space $\mathcal Q_{\text{sh}}$. As we shall soon see, defining a second-order mark-weighted $K$ function requires defining reference measures on $\mathcal X$ and $\mathcal M$. When $\mathcal M$ is a quotient of $\mathcal Q$ containing equivalence classes of curves, a measure on $\mathcal Q$ invariant to the group of symmetries is a natural choice for a reference measure. However, no such measure exists since each symmetry group under consideration is infinite-dimensional, on account of the Lie group $\Gamma$ of reparameterizations of $\mathbb S^1$. This presents significant difficulties in constructing a well-defined $K$ function, and in its estimation. Our strategy is to thus consider $\mathcal M=\mathcal Q$, a Polish space (Section 1 in supplement), with Borel sets $\mathbb B(\mathcal M)$, but introduce the necessary symmetries into the $K$ function via a weight or test function. 

At each point $x_i \in \mathcal X$ we attach a mark $(q_i *g_i) \in \mathcal Q$, resulting in the marked point process $\Psi = \{(x_i, (q_i*g_i))\}_{i=1}^N$ on $(\mathcal X \times \mathcal M, \mathbb B(\mathcal X \times \mathcal M))$. Let the spatial domain $\mathcal X$ be equipped with the Lebesgue measure $\dd x$ on $\mathbb  R^2$ as its reference measure. The reference measure on  $\mathcal M$ is $\nu$, defined as a pushforward of a product reference measure on $\mathcal Q \times G$ under the continuous group action map $(q,g) \mapsto q*g$; see Section 1 in supplement for a precise definition. This results in the product reference measure $\lambda:=\dd x \otimes \nu$ on $\mathbb B(\mathcal X \times \mathcal M)$. The symmetry $g_i$ associated with curve $q_i$ is interpretable only in a relative sense with respect to another curve -- this is the central aspect of establishing correspondence. A consequence of this is seen in our definition of the $K$ function, and its estimation, wherein a template is assumed.

\subsection{Product density and pair correlation}
Assume that the first-order product density $\rho$ and the second-order product density $\rho^{(2)}$ of $\Psi$ exist. Then, 
from the Campbell formula \citep{chiu2013stochastic}, for a measurable function $f:\mathcal X \times \mathcal M \to [0,\infty)$,
\begin{align}
	&\mathbb{E}\left[\sum^{\neq}_{(x_1,q_1 *g_1),(x_2,q_2*g_2)\in\Psi}f((x_1,q_1 *g_1),(x_2,q_2*g_2))\right] \nonumber\\
	\qquad &= \displaystyle \int_{(\mathcal{X}\times G)^2}
	f((x_1,q_1 *g_1),(x_2,q_2*g_2))
	\, \rho^{(2)}((x_1,q_1 *g_1),(x_2,q_2*g_2)) 
\prod_{i=1}^2\dd \lambda(x_i ,q_i *g_i),
	\label{eq:campbell}
\end{align}
where $\neq$ represents summation over distinct tuples, and the measure $\prod_{i=1}^2\dd \lambda(x_i ,q_i *g_i)=\prod_{i=1}^2\dd x(x_i) \dd \nu (q_i*g_i)$. The second-order product density $\rho^{(2)}$ may be expressed in terms of the second-order product density $\rho^{(2)}_{\text{gr}}$ of the ground process $\Psi_{\text{gr}}$ as \citep{heinrich2012asymptotic}
\begin{equation}	\label{eq:prod_density}
\rho^{(2)}((x_1,q_1 *g_1),(x_2,q_2*g_2)) = J_{x_1,x_2}\left(q_1*g_1,q_2*g_2\right)\rho_{\text{gr}}^{(2)}(x_1,x_2), 
\end{equation}
where $J_{x_1,x_2}\left(q_1*g_1,q_2*g_2\right)$ represents the family of conditional joint densities of $(q_1*g_1,  q_2*g_2)$ given pairs of locations $(x_1, x_2)$, where the densities are with respect to the reference measure $\nu$. Indeed, families of conditional densities $J_{x_1,\ldots,x_k}$ exists for every $k \leq N$. 

Denote by $\eta_{\text{gr}}(x_1,x_2)=\frac{\rho^{(2)}_{\text{gr}}(x_1,x_2)}{\rho_{\text{gr}}(x_1)\rho_{\text{gr}}(x_2)}$ the pair correlation function of the ground process $\Psi_{\text{gr}}$. 
Then, a more interpretable second-order measure of dependence within points in $\Psi$ is its \emph{pair correlation function}, defined as
\begin{eqnarray}
	\eta_{\Psi}=\frac{J_{x_1,x_2}}{J_{x_1}J_{x_2}}	\eta_{\text{gr}}.
	\label{eq:pair-correlation-functional}
\end{eqnarray}
In \eqref{eq:pair-correlation-functional}, the first term represents the conditional joint density of the marks, providing insight into mark interactions given spatial locations. Meanwhile, $\eta_{\text{gr}}$ reveals interactions between spatial points: values greater than 1 indicate clustering, values less than 1 suggest repulsion, and a value of 1 represents spatial independence. Although it is not a correlation in the traditional sense, this pair correlation function is convenient for examining second-order dependencies in spatial patterns.

We make the following three assumptions.
\begin{description}
\itemsep -0.5em
			\item[A1.] The pair correlation function $\eta_\Psi$ satisfies 
				$\eta_{\Psi}((x_1,q_1*g_1), (x_2,q_2*g_2)) = \eta_{\Psi}((x_1+v,q_1*g_1), (x_2+v,q_2*g_2))$   for every $v \in \mathcal{X}$. 
	\item[A2.] The densities $J_x$ are the same at all $x \in \mathcal X$, and the mark distribution is hence the same at all locations. 
	\item [A3.]  The mark distribution equals the reference measure $\nu$.  
\end{description}
Under Assumption A1, the process $\Psi$ is said be second-order intensity-weighted stationary \citep{iftimi2019second}, which is a form of weak stationarity of $\Psi$. Further, dependence between points $(x_i,q_i*g_i)$ in $\Psi$ depends only on the relative distances between their spatial locations $x_i$, regardless of the marks. Assumption A1 is consistent with what is required to answer the motivational question (ii) in the Introduction. Assumption A2 offers a simplification when defining the $K$ function. If, additionally, the marks are independent across spatial locations, the situation is referred to as \emph{random labeling} \citep{ghorbani2021functional}. Assumption A3 simplifies computation of the $K$ function, and will be discussed later.

\section{Mark-weighted $K$ function and estimation}
\label{sec:Kfunction}

We first define the population $K$ function that represents a second-order dependence measure of the marked point process $\Psi$. The definition is based on the functional mark-weighted $K$ function introduced by \cite{ghorbani2021functional}, suitably adapted to the setting of parameterized curves in the presence of a symmetry group $G$. 

\subsection{$K$ function}
The mark-weighted $K$ function extends classical $K$ functions by incorporating a test function that quantifies dependency in variations between marks at different locations. We consider three test functions $ \ff : \mathcal Q \times \mathcal Q \to \mathbb R_{\geq 0}$ arising from the three distances $d_{\text{sh}},\ d_{\text{sc-sh}},\ d_{\text{ro-sh}}$ on curves, corresponding to three different symmetry groups, resulting in three $K$ functions, which capture correlations between the spatial locations and nuclei shapes. The $K$ function is obtained by weighting the second-order product densities at points within balls of radius $r>0$ on the spatial domain $\mathcal X$ by the test function. 

Denote by $c_\ff:= \mathbb{E}(\ff(q*g, q_1*g_1) \mid q*g, q_1*g_1 \perp X)$ the expectation of the test function $\ff$ under random labeling, where the marks are independent of the locations. Let $B_x(r)$ be the open ball in $\mathbb R^2$ of radius $r>0$ centered at a point $x$, and let $\mathbb I_A$ denote the indicator function on the set $A$. Under Assumption A1 of stationarity of $\Psi$, a sensible mark-weighted $K$ function may be defined as
\begin{align*}
	K_{\ff}(r)
	&= \frac{1}{|\mathcal X|c_{\ff}} \mathbb{E} \left[ \sum_{(x,q*g) \in \Psi} \sum_{(x_1,q_1*g_1) \in \Psi \setminus \{(x, q*g)\}} \frac{\ff(q*g, q_1*g_1)}{\rho(x, q*g)} \frac{\mathbb I \{x_1 \in \mathcal X \cap B_x(r)\}}{\rho(x_1, q_1*g_1)} \right],
\end{align*}
where $\rho$ is the first-order intensity of $\Psi$ and $|\mathcal X|$ is the area of the spatial domain. The test function $\ff$ summarizes dependency amongst the shape marks, and the $K$ function thus represents the average value of $\ff$ for all pairs of marks separated by a distance $r$, normalized by the first-order product density to remove the effect of spatial intensity variations. The $K$ function reflects how the similarity (or dissimilarity) of the marks depends on spatial proximity, allowing us to detect clustering, dispersion or spatial independence of the marks. 
Under random labeling, the normalizing constant is $c_\ff= \mathbb  E(\|q*g-m_G\|^2)$, where $m_G$ is $\mathbb E(q*g)$ for $g \in G$ with respect to the measure $\nu$ on the mark space (Section 2 in the supplement). It is however difficult to estimate the product densities $\rho(x,q*g)$, especially since the mark space is infinite-dimensional and nonlinear. 

We thus consider a simplified definition of the $K$ function obtained by further making Assumptions A2 and A3: upon assuming a common mark distribution, from \eqref{eq:prod_density} we observe that if the mark distribution equals the reference measure, we get $\rho(x,q*g)=\rho_{\text{gr}}(x)$ for any $q \in \mathcal Q$ and $g \in G$, since the family of conditional densities $J_{x}(q*g)=J(q*g)\equiv 1$ for all $x \in \mathcal X$. Such an approach is also used in \cite{iftimi2019second,ghorbani2021functional, d2024local}. Assumption A3 represents a practical solution given the infinite-dimensionality and non-Euclidean nature of the mark space $\mathcal M$. We thus consider a mark-weighted $K$ function defined as follows. 

\begin{definition}
	\label{def:Kfunction}
Under the Assumptions A1-A3  of the process $\Psi$, the \emph{mark-weighted $K$ function} $r \mapsto K_\ff(r)$, based on test function $\ff$ corresponding to a symmetry group $G$, is defined by
\begin{align}
	K_{\ff}(r)
	&= \frac{1}{|\mathcal X|c_{\ff}} \mathbb{E} \left[ \sum_{(x,q*g) \in \Psi} \sum_{(x_1,q_1*g_1) \in \Psi \setminus \{(x, q*g)\}} \frac{\ff(q*g, q_1*g_1)}{\rho_{\text{gr}}(x)} \frac{\mathbb I \{x_1 \in \mathcal X \cap B_x(r)\}}{\rho_{\text{gr}}(x_1)} \right].
	\label{eq:marked-k}    
\end{align}
\end{definition}

The definition employs scaling by the intensity $\rho_{\text{gr}}$ of the ground process to ensure that, under random labeling, $K_\ff(r)=\pi r^2$, similar to what would be the case if the ground process $\Psi_{\text{gr}}$ were to be a homogeneous Poisson point process on $\mathcal X$. This property of $K_\ff$ under random labeling will be used in Section \ref{sec:hypothesis-testing} when testing for the spatial dependence of the shape marks. We use three test functions $\ff$ based on the three curve distances:
$$\ff(q*g,q_1*g_1):=\frac{1}{2}d_j(q*\hat g, q_1*\hat g_1)^2, \quad j \in \{\text{sh},\text{sc-sh},\text{ro-sh}\},$$
where $\hat g$ and $\hat g_1$ are the symmetries of $q$ and $q_1$ that are in optimal correspondence with a template curve, say, $\mu_{\text{temp}}$; our definition of the mark-weighted $K$ function implicitly then depends on the template $\mu_{\text{temp}}$. The need for a template is due to the fact that the symmetries are interpretable only in a relative sense, and, moreover, is unavoidable in practice when working with quotient spaces under group actions. The factor of $\frac{1}{2}$ in $\ff$ cancels out a factor of 2 that arises in the derivation of the normalizing factor $c_\ff$ (Section 2 in the supplement). Finally, the distance-based test functions may be interpreted as variograms that account for different symmetry groups.

\subsection{Estimator for the $K$ function and its computation}
Given curves $q_1,\ldots,q_N$, to estimate the $K$ function in \eqref{eq:marked-k}, we need to perform the following tasks: 
(i) choose a fixed template $\mu_{\text{temp}}$ or estimate one, (ii) estimate the unobserved shape symmetries $g$ using the template ((i) and (ii) together help evaluate the test function $\ff$), 
(iii) estimate the normalizing constant $c_\ff$,
(iv) estimate the intensity $\rho_{\text{gr}}$ of the ground process $\Psi_{\text{gr}}$, and
(v) correct for the boundary effect of $\mathcal X \subset \mathbb R^2$.

Two estimators of $K_\ff$ may be considered. If the symmetries $g$, constant $c_\ff$ and intensity $\rho_{\text{gr}}$ are \emph{known} or may be recovered exactly, then the estimator
\begin{equation}
\bar {K}_{\ff}(r) =
\frac{1}{|\mathcal X|c_{\ff}}  \sum_{(x,q*g) \in \Psi} \sum_{(x_1,q_1*g_1) \in \Psi \setminus \{(x, q*g)\}} 
\frac{w(x,x_1)\ff(q*g, q_1*g_1)}{\rho_{\text{gr}}(x)\rho_{\text{gr}}(x_1)} \mathbb I_{x \in \mathcal X}\mathbb I \{x_1 \in \mathcal X \cap B_x(r)\},
\label{eq:estimator_K}     
\end{equation}
where $w$ is an edge correction function satisfying $\int_{\mathcal X} \mathbb I\{(x+r) \in \mathcal X\} \, w(x,x+r) \, \dd x = |\mathcal X|$, is unbiased for $K_\ff(r)$ for every $r>0$; see Section 3 in the supplement for a proof. On the other hand if the symmetries are estimated consistently as sample size increases within $\mathcal X$ (infill asymptotics),  the estimator $\hat K$ defined as $\bar K_\ff$ with $g$, ground process intensity $\rho_g$ and normalizing constant $c_\ff$ replaced by estimators $\hat g, \hat g_1, \hat \rho_{\text{gr}}, \hat \ff$ and $\hat c_\ff$ in place of $g, g_1,\rho_{\text{gr}}, \ff$ and $c_\ff$, respectively, can be shown to be asymptotically unbiased with growing sample size within $\mathcal X$; see Section 4 in the supplement for a discussion on consistency of $\hat K_{\ff}$. 

Examples of edge correction functions $w$ satisfying the above requirement include the minus sampling edge correction $w_{\ominus}(x,x+y)=\frac{|\mathcal X|\mathbb I\{x\in \mathcal X\ominus B_{0}(|y|)\}}{|\mathcal X\ominus B_{0}(|y|)|}$, where $\ominus$ denotes Minkowski subtraction ($A\ominus B=\{x:x+A\subseteq B\}$), the translational edge correction $w_{\cap}(x,x+y)=\frac{|\mathcal X|}{|(\mathcal X+(x+y))\cap (\mathcal X+x)|}$, and the isotropic edge correction 
$w\bigl(x,x+y\bigr) = \frac{2\pi \,|y|}{l\bigl(\partial B_x(|y|)\,\cap\,\mathcal X\bigr)}$, 
where $l$ is the length of the curve representing the boundary. Note that by an abuse of notation $|\cdot|$ denotes the area of a spatial domain as well as the Euclidean norm of a vector in $\mathbb{R}^2$.

In practice, however, the symmetries $g$, constant $c_\ff$ and intensity $\rho_{\text{gr}}$ are unknown, and need to be estimated from the data. At a location $x$ with mark $q*g$, for every $x_1$ with mark $q_1*g_1$ within the ball of radius $r$ around $x$, the symmetries $g$ and $g_1$ are estimated by aligning, or establishing correspondence between, $q$ and each $q_1$ with respect to the template. For the chosen test function $\ff$ based on the distances $d_j,\ j \in \{\text{sh},\text{sc-sh},\text{ro-sh}\}$, we propose using the corresponding Karcher mean from Section \ref{subsec:mean_geo} as an estimate of the template $\mu_{\text{temp}}$. Thus, as described in Section \ref{subsec:mean_geo}, we compute the Karcher mean and perform joint alignment of $q_1, \ldots,q_N$ to the Karcher mean using an iterative procedure, resulting in the following estimates: (i) $\hat g,\ \hat g_i$ for $i=1,\ldots,N$, (ii) Karcher mean $\mu_{j},\ j \in \{\text{sh},\text{sc-sh},\text{ro-sh}\}$ as estimate of the template $\mu_{\text{temp}}$ depending on the choice of the test function $\ff$, (iii) upon combining these, estimate $\hat \ff(q*\hat g,q_1*\hat g_1)$ of the value of $\ff$. 

Since the normalizing constant $c_\ff=\mathbb  E(\|q*g-m_G\|^2)$, where $m_G$ is $\mathbb E(q*g)$ for $g \in G$, we consider the estimate
$\hat{c}_{\ff} = \frac{1}{N} \sum_{i=1}^N\|(q_i*\hat g_i)-\mu_j\|^2,\ j \in \{\text{sh},\text{sc-sh},\text{ro-sh}\}$, where $\mu_j$ is determined by the choice of $\ff$. Finally, we estimate the intensity $\rho_{\text{gr}}$ of the ground process $\Psi_{\text{gr}}$ using a Gaussian smoothing kernel. In later sections, we use the \texttt{density.ppp} function in the \texttt{spatstat} \texttt{R} package, where the bandwidth is selected following the approach of \cite{cronie2018non} using the \texttt{bw.CvL} function in the same package. 

\section{Testing for spatial dependence of nuclei shapes}
\label{sec:hypothesis-testing}
The mark-weighted $K$ function and its estimate $\hat K$ may be used to test if the nuclei shapes correlate with their spatial locations by comparing $\hat K$ to the value of $K$ under the null hypothesis of random labeling; under the null hypothesis, $K(r)=\pi r^2$ regardless of the intensity $\rho_{\text{gr}}$ of the ground process $\Psi_{\text{gr}}$. This ensures that available tests for point processes, not necessarily with marks, may in principle be used. 

The pipeline from a histopathology image to a realization $\Psi$ of the marked point process is noisy. It hence may be the case that spatial correlations between nuclei shapes manifest at different scales characterized by the radius $r$. We thus consider two types of tests: (i) a local envelope test \citep{ripley1977modelling}, which presents evidence for correlations for each radius $r$, and (ii) a global envelope test \citep{myllymaki2017global}, which provides evidence for correlations uniformly over all radii $r$. These tests were originally designed for point processes without marks. Adaptation to accommodate shape marks is straightforward owing to the aforementioned property of the definition of our $K$ function.

The test statistic we use is based on a transformed version $\hat L(r)=\sqrt{\frac{\hat K(r)}{\pi}}$ of $\hat K$: 
\begin{equation}
	T(r)=\hat L(r) - L_{H_0}(r),
	\label{eq:centeredL}
\end{equation}
where $L_{H_0}=r$ if we wish to test for random labeling. The distribution of $T(r)$ is approximated via resampling under the null hypothesis of independence between marks and their locations, by permuting the marks while keeping their locations fixed. The test statistic with $L_{H_0}=r$ would result in a fairly conservative rejection region, since $L_{H_0}=r$ is sensible only when unknown components (e.g., symmetries, ground intensity $\rho_{\text{gr}}$) of the population $K$ function in \eqref{eq:marked-k} are known \emph{a priori}. Instead, as recommended by \cite{grabarnik2011correct}, we choose $L_{H_0}$ as the $\hat L$ obtained from the ground process when the test function $\ff(\cdot,\cdot)\equiv 1$. 

Once multiple samples of $T(r)$ have been obtained through resampling, for the local and global envelope tests, we provide, respectively:
\begin{itemize}
	\itemsep -0.5em
	\item the proportion of radii for which the test statistic deviated from the 95\% pointwise (in $r$) envelope generated using $s=2499$ permutations \citep{ripley1977modelling};
	\item the p-value derived from the global rank envelope test using the \emph{extreme rank depth} measure on the samples with $s=2499$ permutations \citep{myllymaki2017global}.
\end{itemize}
The global envelope test is implemented using the  \texttt{global\_envelope\_test} function in the package \texttt{GET} in \texttt{R}. By carrying out the two types of tests, we note that if spatial dependence is asserted by the global test, further probing of dependence structures at local scales based on different values of $r$ may be carried out by using the local test, along the lines of the method recently proposed by \cite{d2024local}. 

\section{Simulations}
\label{sec:simul}

We evaluate the effectiveness in detecting spatial dependence among shape, size-and-shape or orientation-and-shape marks based on the proposed mark-weighted $K$ functions using simulations. In all cases, spatial dependence among marks is imposed via inter-point distances. We first simulate realizations of a homogeneous Poisson process (PP) over the domain $[0,4] \times [0,4]$ with an intensity of $\rho_{\text{gr}} = 8$, resulting in 128 locations on average. Computing for all experiments was performed on a node with 48 cores (Intel(R) Xeon(R) Platinum 8260 CPU $@$ 2.40GHz) with 187GB of memory. Estimation of a single shape mark-weighted $K$ function in the simulations requires approximately $25$ minutes on average (across ten replicates); no parallel computing was used. The computing time depends on (i) the number of curves in the spatial region of interest (average of 131), and (ii) discretization fineness for each curve (100).

\subsection{Simulation of marks} 

We generate spatially correlated shapes at the sampled locations using the Fourier basis functions $\{\cos(lt),\ \sin(lt),\ l=0,1,2,\ t\in[-\pi,\pi]\}$. Letting $\{b_j,\ j=1,\dots,6\}$ denote the basis functions, the radial distance as a function of the angle for curve $i$ is defined as 
\begin{equation}
    \Gamma_i(t) = \sum_{j=1}^{6} c_{ij}b_j(t),
    \label{eq:fourier-basis-func}
\end{equation} 
where $\{c_{ij},\ i=1,\dots,N,\ j=1,\dots,6\}$ are Fourier coefficients. 
Each shape is then obtained by transforming the radial distance function to Cartesian coordinates: $(x_i(t)=\Gamma_i'(t)\cos(t),\ y_i(t)=\Gamma_i'(t) \sin(t))$,
where $\Gamma_i'(t)=\Gamma_i(t)+|\underset{t}{\min}\Gamma_i(t)|+|\underset{t}{\max}\Gamma_i(t)|
$ is an adjusted radial distance function that ensures that the resulting shapes are non-self-intersecting. 

Spatial dependence in shape, rotation and scale is introduced independently. Shape dependence is incorporated through the Fourier coefficients $c_{ij}$, which are sampled from $\text{MVN}(0_N^\top, C_{\text{mat}})$ to induce spatial dependence, and from $\text{MVN}(0_N^\top, 0.1I_N + 0.9J_N)$ for the independent case. Here, MVN is short for the multivariate Normal distribution, $0_N^\top$ is an $N$-dimensional vector of zeros, $C_{\text{mat}}$ is the Mat\'{e}rn covariance matrix with scale parameter of 0.5, smoothness parameter of 0.5 and range parameter of 2, $I_N$ is an $N\times N$ identity matrix, and $J_N$ is an $N \times N$ matrix of ones. To induce dependence in the shapes' orientations, we generate two unit vectors $e_{i1}$ and $e_{i2}$ independently from $\text{MVN}(0_N^\top, C_{\text{mat}})$, and compute the rotation angle $\phi_i$ between them; as before, $C_{\text{mat}}$ is the Mat\'{e}rn covariance with the same parameter values. In the independent case, $\phi_i$ is sampled independently from $\text{Unif}(0,2\pi)$. To model scale dependence, the scale factor $\sigma_i$ is sampled from the correlated uniform distribution on $[0.2,1.2]$, by transforming a sample from $\text{MVN}(1_N^\top, C_{\text{mat}})$. For the independent case, the Mat\'{e}rn covariance is replaced by the identity matrix $I_N$. The resulting shapes, with additional rotations and rescalings, are given by $\beta_i(t)=(x_i^*(t) = \sigma_i[x_i(t)\cos(\phi_i)-y_i(t)\sin(\phi_i)],\ y_i^*(t) = \sigma_i[x_i(t)\sin(\phi_i)+y_i(t)\cos(\phi_i)])$.

\begin{figure}[!t]
    \centering
    \renewcommand{\arraystretch}{0}
    \begin{tabular}{@{}c@{}c@{}c@{}}
         \includegraphics[width=0.33\textwidth]{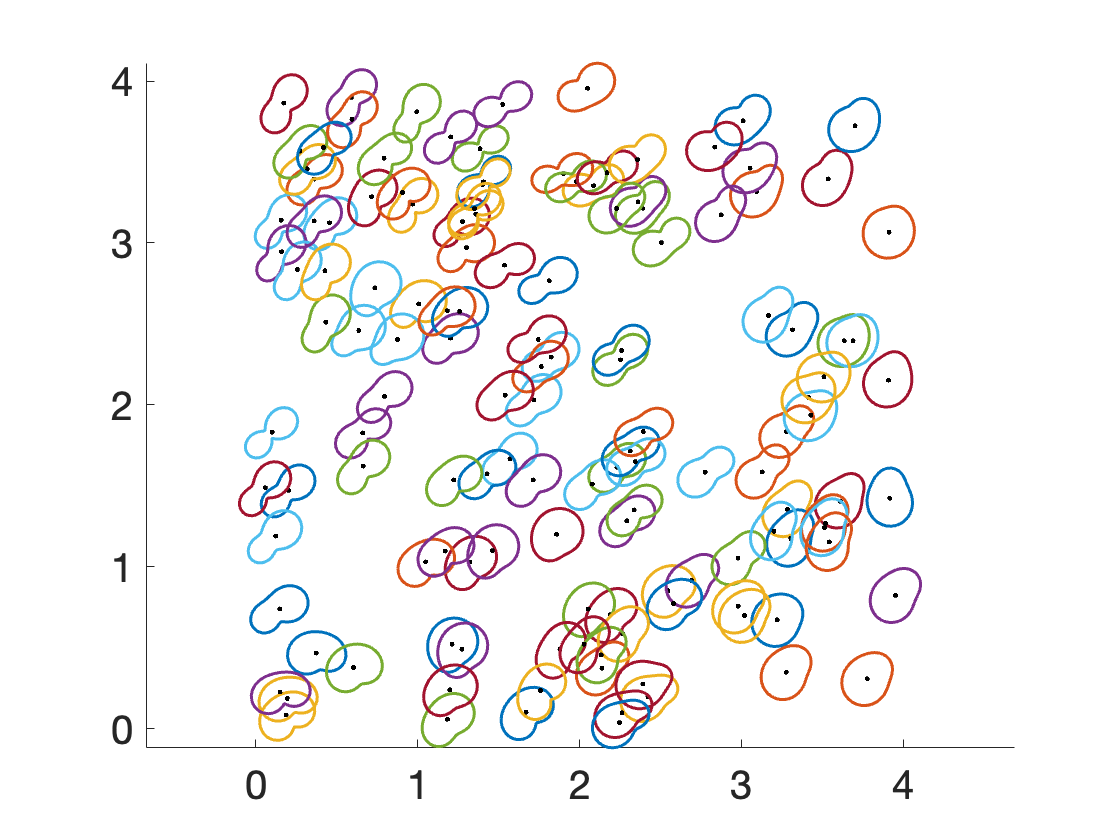}
         &\includegraphics[width=0.33\textwidth]{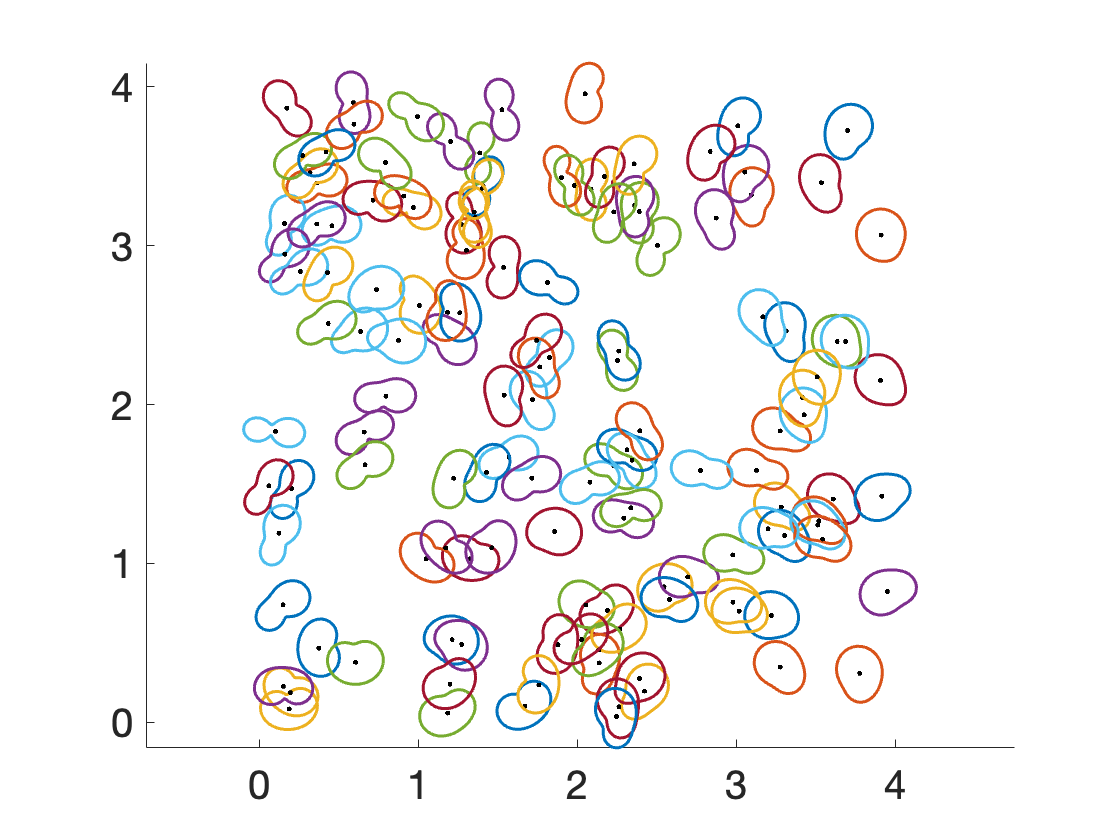}
         &\includegraphics[width=0.33\textwidth]{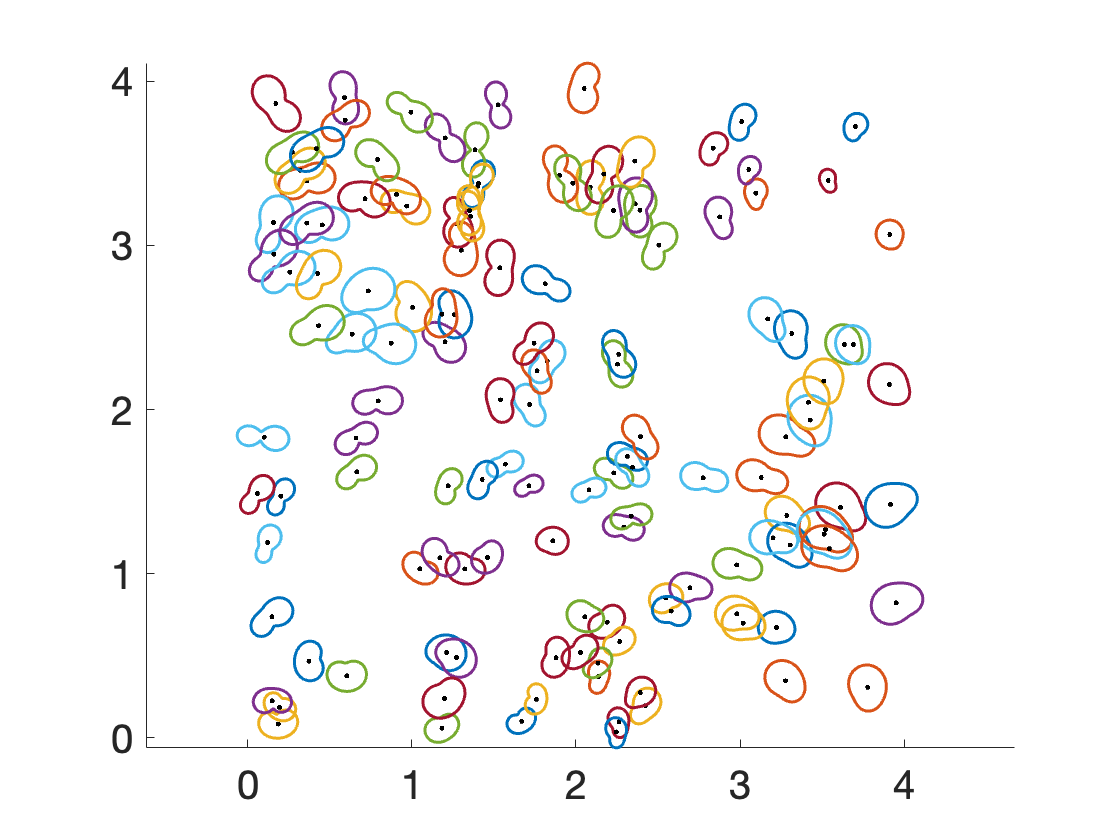}\\
         (a)&(b)&(c)
    \end{tabular}
    \caption{\small Simulated (a) spatially dependent shapes, (b) the same shapes as in (a) with spatially dependent orientations, and (c) the same shapes as in (b) with spatially dependent sizes.}
    \label{fig:simul-shape}
\end{figure}

Figure~\ref{fig:simul-shape} illustrates the sequence of steps taken to generate one realization of different types of marked point processes. Panel (a) shows a shape marked point process, where nearby curves tend to have similar shapes. Panel (b) presents the same curves as in (a), but with additional spatially dependent orientations. Panel (c) shows the same curves as in (b), but after additional spatially dependent rescalings.

\subsection{Results}

We evaluated eight scenarios, varying spatial dependence in shape, orientation and size. We computed the proposed mark-weighted $K$ functions using the shape, orientation-and-shape and size-and-shape test functions and performed global and local tests for each scenario. The results, based on 50 independent replicates, are summarized in Table \ref{tab:simu-shape-prop}. Columns in (a) report the spatial dependence structure in the (i) shape, (ii) orientation and (iii) size of the marks, with `0' corresponding to independence and `1' to dependence. Columns in (b) report the average p-values from the global rank envelope tests (using extreme rank depth), at a significance level of $\alpha=0.05$, based on the (i) shape, (ii) orientation-and-shape and (iii) size-and-shape mark-weighted $K$ functions. Similarly, columns in (c) report average proportions of radii $r$ for which the (i) shape, (ii) orientation-and-shape and (iii) size-and-shape mark-weighted $K$ functions deviated from the pointwise 95$\%$ envelope. In both cases, standard deviations are presented in parentheses. We describe the results obtained from the global tests in detail and note that the pointwise tests resulted in very similar patterns.

In the independent mark scenario (shape=orientation=size=0), the global rank envelope tests based on all three mark-weighted $K$ functions resulted in non-significant and fairly high p-values. On the other hand, when the marks' shape, orientation and size were all spatially dependent (shape=orientation=size=1), all three global tests resulted in extremely small p-values; this result is reassuring as it represents the scenario with the strongest spatial dependence signal.

\begin{table}[!t]
\centering 
\renewcommand{\arraystretch}{1}
\setlength{\tabcolsep}{3pt}
\begin{tabular}{|c|c|c||c|c|c||c|c|c|}
\hline
\multicolumn{3}{|c||}{\textbf{(a)}} & \multicolumn{3}{c||}{\textbf{(b)}} & \multicolumn{3}{c|}{\textbf{(c)}} \\
\hline
(i) & (ii) & (iii) & (i) & (ii) & (iii) & (i) & (ii) & (iii) \\
\hline
0 & 0 & 0 & 0.52 (0.31) & 0.48 (0.30) & 0.46 (0.28) & 0.04 (0.10) & 0.05 (0.08) & 0.05 (0.09) \\
1 & 0 & 0 & 0.00 (0.00) & 0.15 (0.16) & 0.12 (0.22) & 0.98 (0.03) & 0.26 (0.23) & 0.50 (0.31) \\
0 & 1 & 0 & 0.54 (0.25) & 0.02 (0.10) & 0.47 (0.29) & 0.03 (0.08) & 0.93 (0.19) & 0.05 (0.09) \\
0 & 0 & 1 & 0.52 (0.31) & 0.48 (0.30) & 0.00 (0.01) & 0.04 (0.10) & 0.05 (0.08) & 0.96 (0.08) \\
1 & 1 & 0 & 0.00 (0.00) & 0.00 (0.00) & 0.13 (0.18) & 0.99 (0.02) & 0.96 (0.12) & 0.43 (0.34) \\
1 & 0 & 1 & 0.00 (0.00) & 0.15 (0.16) & 0.00 (0.00) & 0.98 (0.03) & 0.26 (0.23) & 0.99 (0.02) \\
0 & 1 & 1 & 0.54 (0.25) & 0.02 (0.10) & 0.00 (0.00) & 0.03 (0.08) & 0.93 (0.19) & 0.95 (0.06) \\
1 & 1 & 1 & 0.00 (0.00) & 0.00 (0.00) & 0.00 (0.00) & 0.99 (0.02) & 0.96 (0.12) & 1.00 (0.01) \\
\hline
\end{tabular}
\caption{\small (a) Spatial dependence structure: `0'=spatial independence, `1'=spatial dependence; (i) shape, (ii) orientation, (iii) size. Results, across 50 replicates, of (b) global rank envelope tests with $\alpha=0.05$ (average p-values), and (c) local envelope tests (average proportions of radii for which test statistics deviated from pointwise $95\%$ envelope): (i) shape, (ii) orientation-and-shape, (iii) size-and-shape; standard deviations reported in parentheses.}
\label{tab:simu-shape-prop}
\end{table}

When only the marks' shape exhibited spatial dependence (shape=1, orientation=size=0), the shape mark-weighted $K$ function was able to detect this structure: the global test resulted in near zero p-values across all replicates. In contrast, global tests based on orientation-and-shape and size-and-shape resulted in non-significant p-values on average (0.15 and 0.12, respectively). These tests jointly assess variation in both shape and either orientation or size, and, since the marks' orientations and sizes were spatially independent in this setting, their inclusion diluted the signal, reducing the power to detect shape dependence alone. When only size exhibited spatial dependence (shape=orientation=0, size=1), only the size-and-shape-based global test resulted in p-values near zero. While the size-and-shape test function assesses variation in both shape and scale, it seems to have primarily captured scale dependence; the shape- and orientation-and-shape-based tests resulted in non-significant and fairly large p-values. Finally, when only the marks' orientations were spatially dependent (shape=size=0, orientation=1), the shape- and size-and-shape-based global tests resulted in large p-values of $0.54$ and $0.47$ on average, respectively. On the other hand, the test based on the orientation-and-shape mark-weighted $K$ function effectively captured the spatial dependence in the marks' orientations, with an average p-value of $0.02$. However, the standard deviation in this case was relatively large: $0.10$. The relative nature of orientation with respect to shape may have contributed to the slightly larger average p-value and much larger standard deviation as compared to cases of spatially dependent shape ($0.00\ (0.00)$) or size ($0.00\ (0.01)$). To verify this, we conducted an additional study. Initially, the Fourier coefficients $c_{ij}$ in \eqref{eq:fourier-basis-func} were sampled from a multivariate normal distribution with the covariance matrix $(1-\tau)I_N + \tau J_N$ with $\tau=0.9$. As the coefficient on $J_N$ decreases, the $c_{ij}$ become more variable, increasing (independent) shape variation. Repeating our analysis using the orientation-and-shape mark-weighted $K$ function with the coefficient reduced to $0.8$, $0.7$ and $0.6$, the average p-values increased to $0.05\ (0.17)$, $0.07\ (0.19)$ and $0.08\ (0.22)$, respectively. Similarly, the average proportion of radii $r$ for which the mark-weighted $K$ function deviated from the pointwise envelope decreased from $0.93$ to $0.85$, $0.80$ and $0.74$, respectively These results confirm that accuracy of estimating relative orientations is influenced by the extent of shape heterogeneity in the sample.

When two out of three sources of variation, i.e., shape and orientation, shape and size or orientation and size, were spatially dependent, the results of global tests based on the different mark-weighted $K$ functions were consistent. For example, when both the shape and size of marks were spatially dependent, the global tests based on shape or size-and-shape mark-weighted $K$ functions resulted in near zero p-values, while the one based on orientation-and-shape resulted in a non-significant p-value of 0.15. Overall, the presented simulation results demonstrate that the mark-weighted $K$ function, when used with suitable test functions, effectively captures various forms of spatial dependence, allowing for assessment of random labeling in curves with symmetries.

\section{Application to breast cancer data}
\label{sec:real-data}

\subsection{Dataset and preliminary analysis}
\begin{figure}[!htb]
    \centering
    \begin{tabular}{@{}c@{}c@{}c@{}}
         \includegraphics[width=0.31\linewidth]{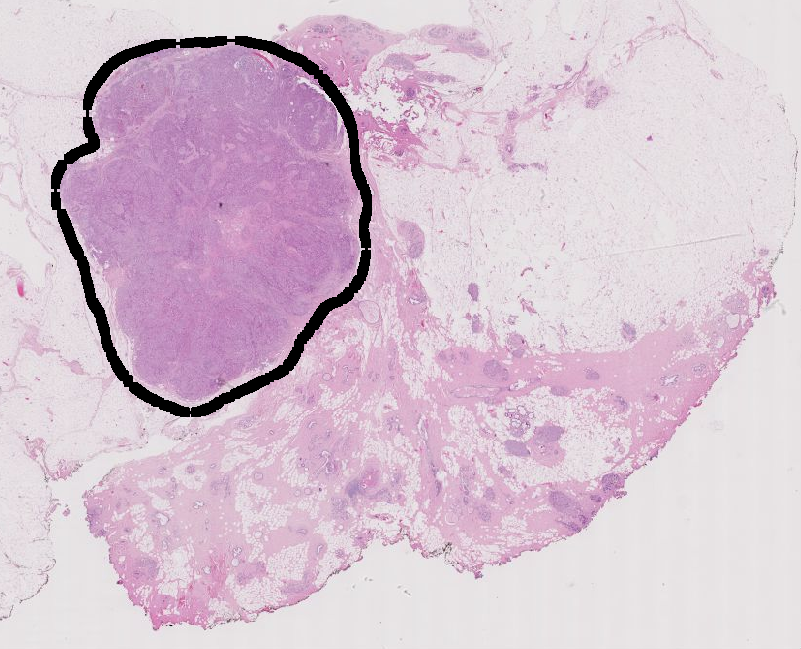} & \includegraphics[width=0.31\linewidth]{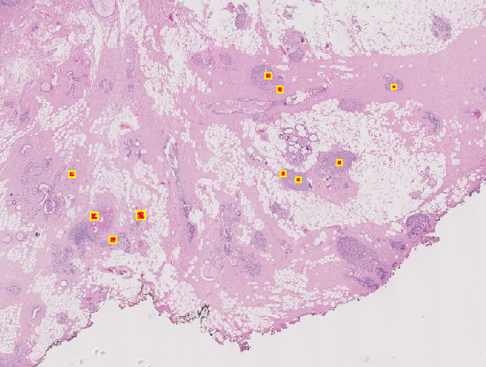} &
         \includegraphics[width=0.31\linewidth]{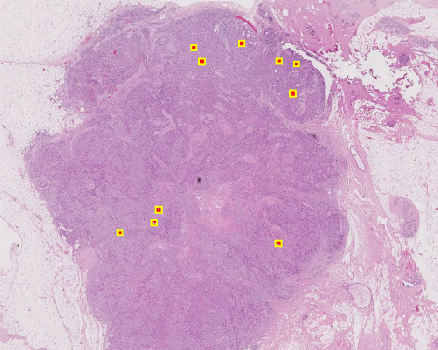} \\
         (a)&(b)&(c)
    \end{tabular}
   \caption{\small (a) Annotated tissue image: tumor area is marked by the black boundary. Ten regions (yellow squares) were randomly sampled within the (b) non-tumor area and (c) tumor area. The red (tumor) and blue (non-tumor) rectangles highlight the areas where regions were sampled.}
    \label{fig:breast-cancer-tissue}
\end{figure}

We examine two types of breast cancer known for poor prognosis: Human Epidermal Growth Factor Receptor 2-positive (HER2+) and Triple Negative (TNBC). We examined five tissue images with manually annotated tumor areas, reviewed by a panel of board certified breast pathologists, from the \cite{tigerChallenge} (Tumor Infiltrating Lymphocytes in Breast Cancer). For each tissue image, we randomly selected ten subregions within both tumor and non-tumor areas, using a fixed square window of $22500\mu m^2$. This resulted in a total of 50 subregions from tumor tissues and 50 subregions from non-tumor tissues. A representative annotated tissue image and the sampled subregions (yellow squares) selected for our analysis are shown in Figure \ref{fig:breast-cancer-tissue}. The tumor is outlined in black. The other four images are displayed in Section 5 in the supplement. The red and blue rectangles highlight the areas where the subregions were sampled. Nuclei outlines were extracted from each subregion using Qupath software \citep{bankhead2017qupath}. Estimation of a single shape mark-weighted $K$ function based on nuclei in a tumor region requires approximately 39 minutes on average (across ten regions; average number of curves per region: 185; discretization fineness: 100). Similarly, based on nuclei in a non-tumor region, estimation requires approximately 29 minutes on average (across ten regions; average number of curves: 134; discretization: 100); as in the simulations, no parallel computing was used.

\begin{figure}[!t]
    \centering
    \begin{tabular}{|c|c|}
    \hline
    \multirow{6}{*}{\includegraphics[width=.2\linewidth]{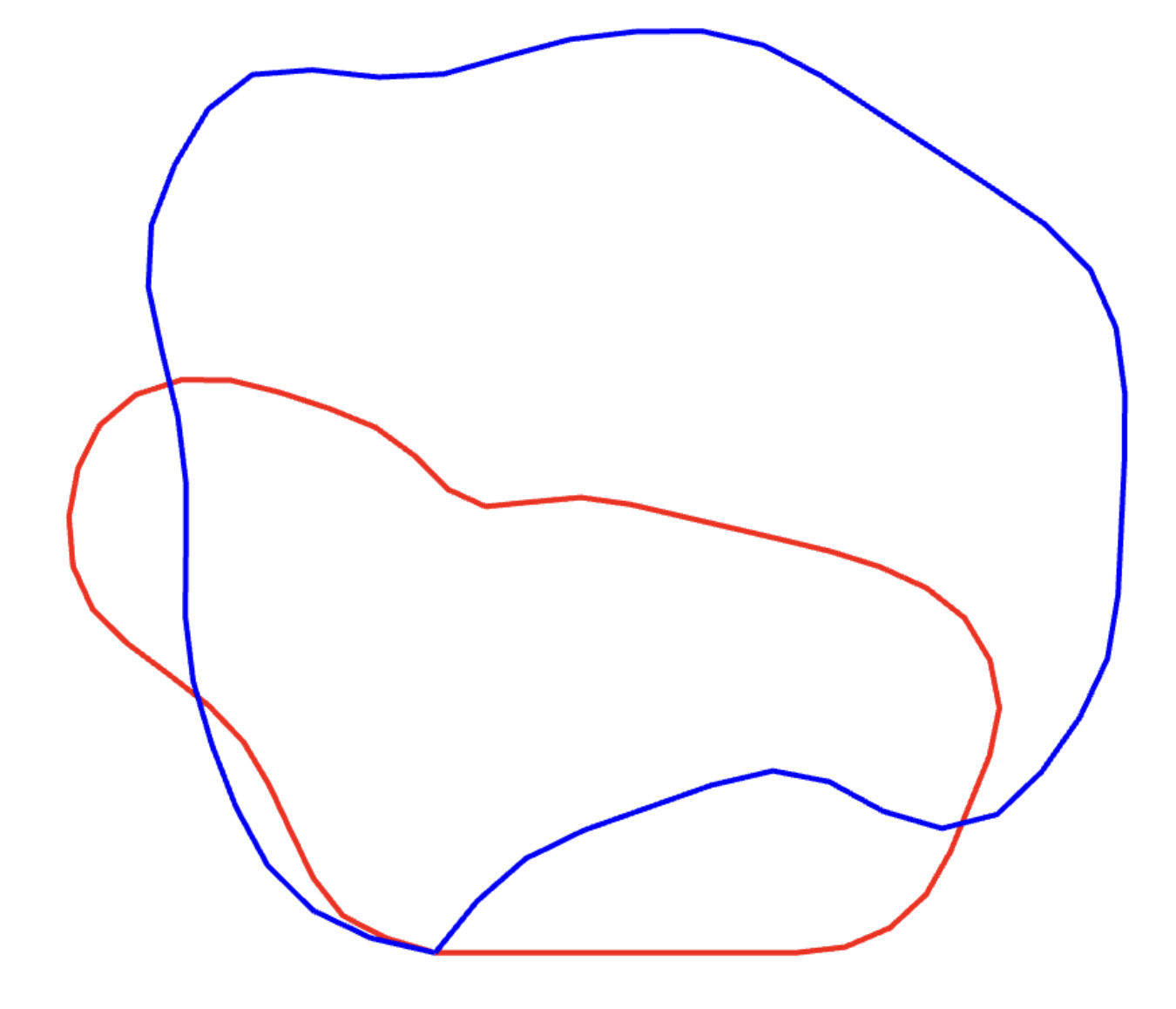}}
    &\includegraphics[width=0.55\linewidth]{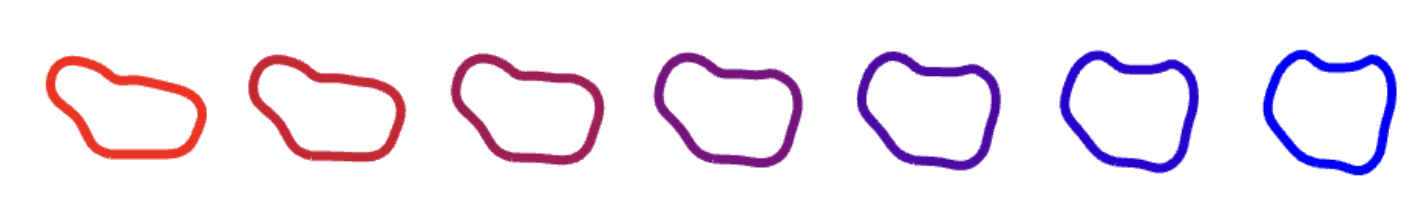}\\
    &(a) $d_{\text{sh}}(q_1,q_2) = 0.3079$\\
    &\includegraphics[width=.55\linewidth]{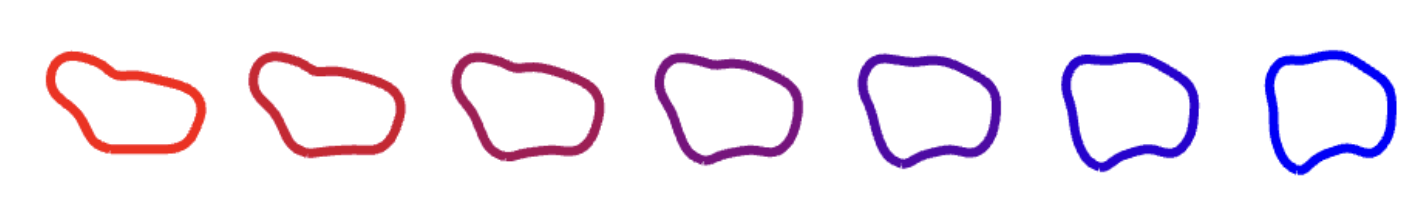}\\
    &(b) $d_{\text{ro-sh}}(q_1,q_2) = 0.3763$\\
    &\includegraphics[width=.55\linewidth]{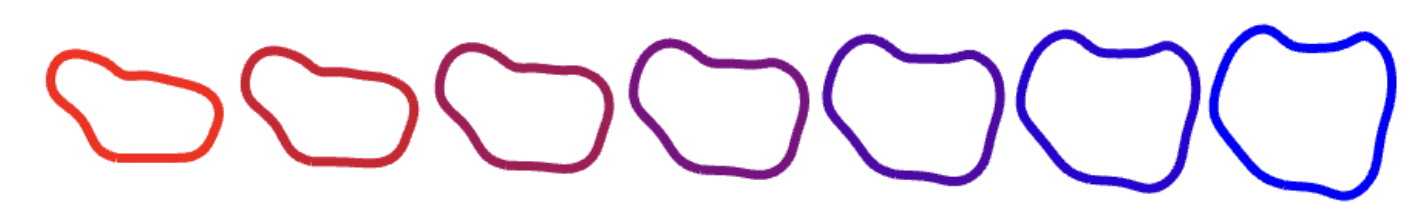}\\
    &(c) $d_{\text{sc-sh}}(q_1,q_2) = 2.5286$\\
    \hline
    \end{tabular}
    \caption{\small Left: Two nuclei boundaries from tumor area. Right: Geodesic paths on the (b) shape, (c) orientation-and-shape, and (d) size-and-shape spaces, with corresponding distances.}
    \label{fig:geodesic}
\end{figure}

We first examine the benefits of accounting for symmetries in cell nuclei outlines by computing geodesics between them as well as their Karcher means, as described in Section \ref{sec:elastic-shape-analysis}. In particular, we assess the symmetries in the cell nuclei as captured by the three distances $d_j,\ j \in \{\text{sh},\text{sc-sh},\text{ro-sh}\}$. Figure \ref{fig:geodesic} compares two tumor cell nuclei via geodesic paths and distances. The two nuclei differ in shape, size and relative orientation, as depicted in the left panel. Panels (a)-(c) illustrate comparisons from different perspectives. In (a), the geodesic path between the two nuclei outlines considers shape only (removing rotation and parameterization variations). Since rotation and parameterization are treated as symmetries, the shape distance is computed after the blue nucleus is optimally rotated and reparameterized with respect to the red one, and both shapes are scaled to have unit length. When orientation is included in the analysis, as shown in (b), the blue nucleus is optimally reparameterized with respect to the red one while maintaining its original orientation. In this case, both shapes are still scaled to unit length since size is also a symmetry. In contrast, panel (c) shows the size-and-shape geodesic wherein the blue nucleus was again optimally rotated and reparameterized with respect to the red one. However, the original sizes of the blue and red nuclei are preserved. When comparing the two nuclei, the geodesic distances vary depending on which factors are accounted for in the analysis. The shape distance between the two nuclei is 0.3079 (panel (a)). In (b), the distance, which now considers both orientation and shape, increases to 0.3763. This larger distance reflects the additional deformation required to account for the natural orientation of the shapes. Lastly, the size-and-shape distance in (c) is larger (2.5286) due to the difference in the natural sizes of the two shapes. 

\begin{figure}[!t]
    \centering
    \begin{tabular}{|c|c|c|}
    \hline
         \multicolumn{3}{|c|}{\includegraphics[width=0.85\linewidth]{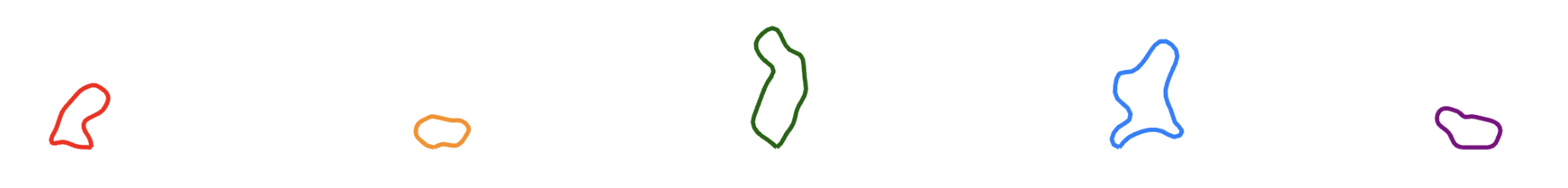}}\\
         \multicolumn{3}{|c|}{(a)}\\
         \hline
         \includegraphics[width=0.27\linewidth]{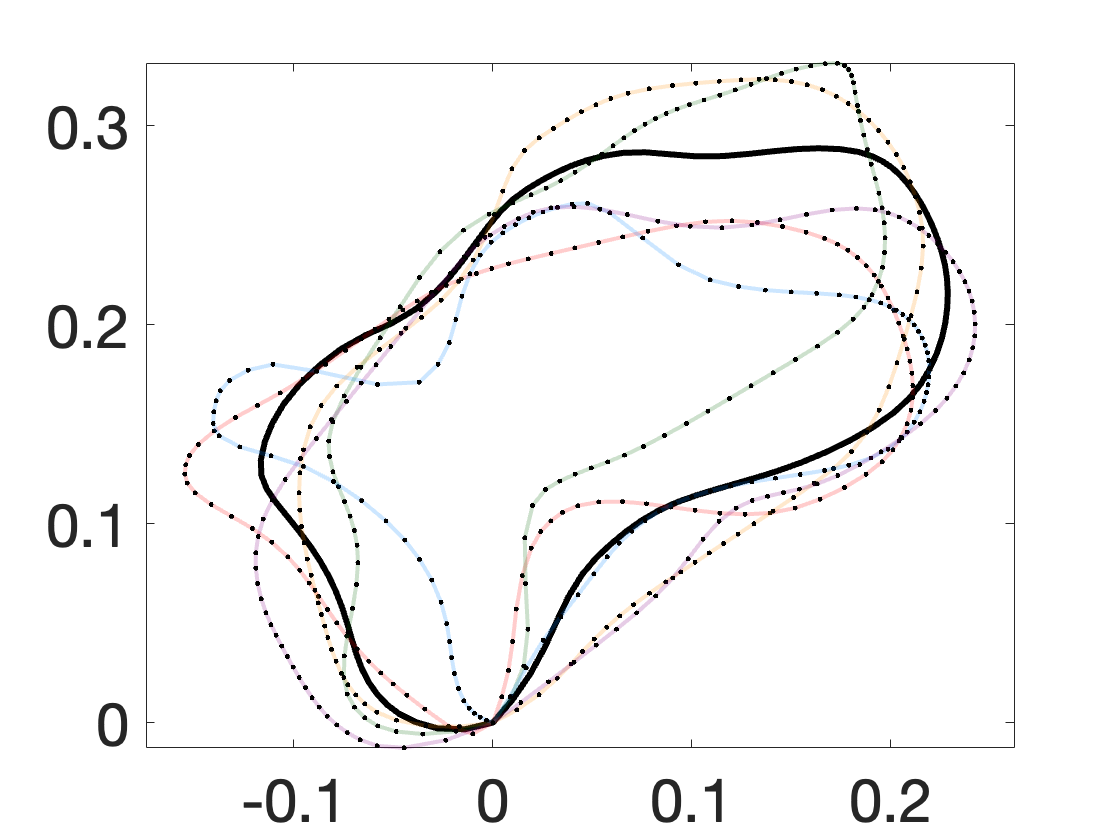}& \includegraphics[width=0.27\linewidth]{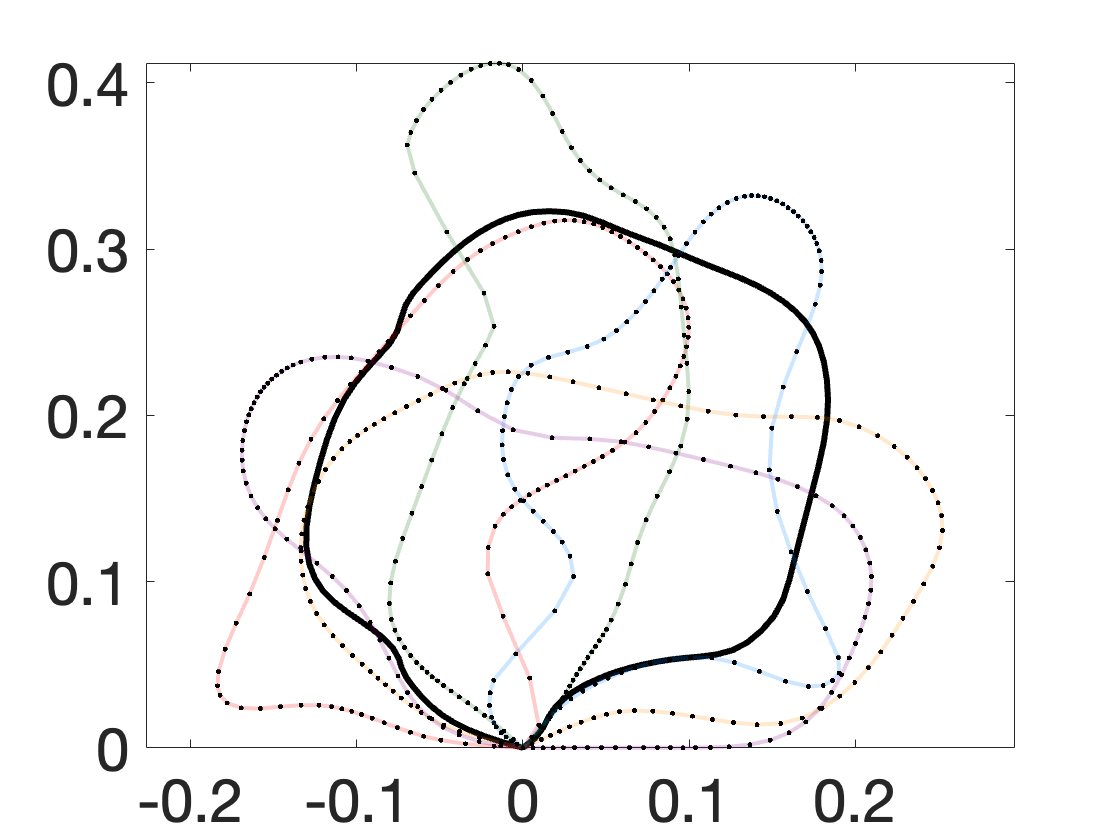}&\includegraphics[width=0.27\linewidth]{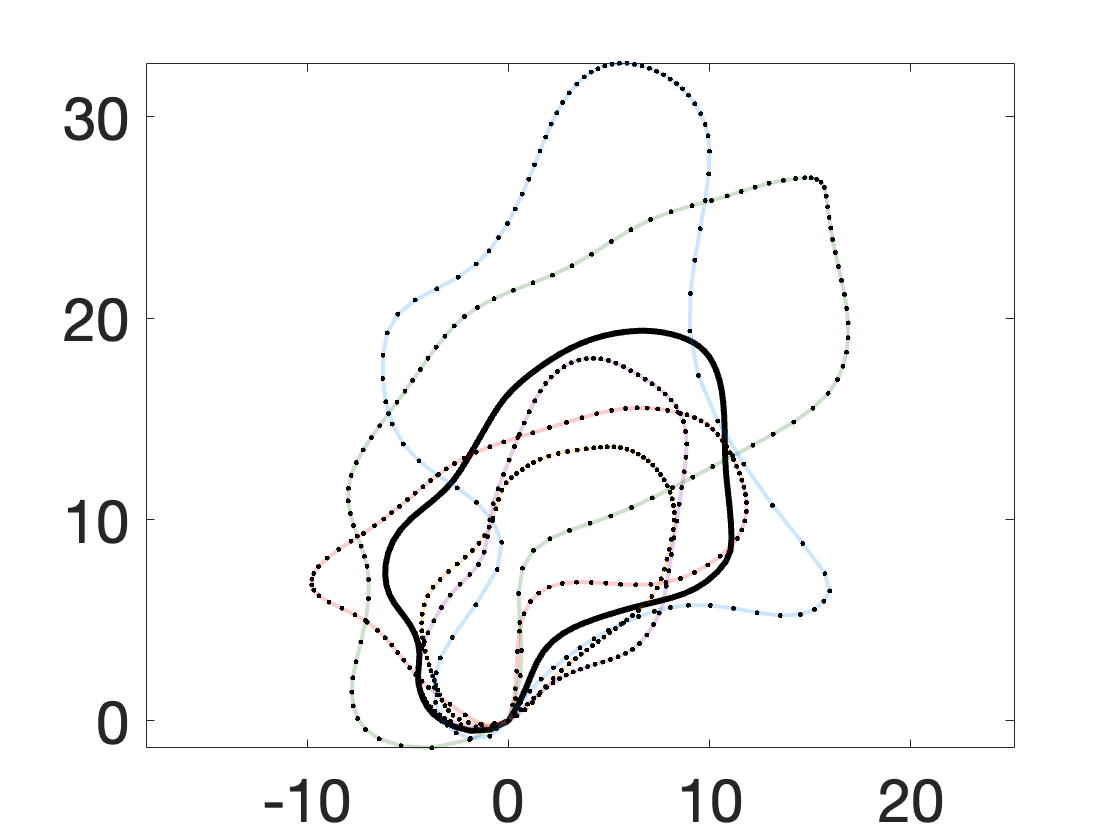}\\
         (b)&(c)&(d)\\
    \hline
    \end{tabular}
    \caption{\small (a) Five nuclei outlines from tumor area. Karcher means (solid black) along with aligned sample nuclei (same colors as in (a)) based on (b) shape, (c) orientation-and-shape and (d) size-and-shape.}
    \label{fig:karcher-mean-shape}
\end{figure}

 Figure \ref{fig:karcher-mean-shape}(a) shows five tumor cell nuclei. Their resulting Karcher means as well as the optimally aligned shapes are shown in panels (b)-(d). In (d), the Karcher mean size-and-shape and all samples retain their natural sizes, as reflected by the axis values, while size variation in (b) and (c) was removed. In both panels (b) and (d), the nuclei are aligned to their Karcher means by optimizing over both rotations and reparameterizations. In contrast, in panel (c), the nuclei retain their natural orientation, with alignment achieved solely by optimizing over reparameterizations. Overall, the results presented in Figures \ref{fig:geodesic} and \ref{fig:karcher-mean-shape} show how the mark-weighted $K$ function captures cell nuclei shape variation. First, the symmetries are accounted for through alignment of the nuclei to an appropriate Karcher mean. Then, variation is captured via appropriate test functions defined using the three distances $d_j,\ j \in \{\text{sh},\text{sc-sh},\text{ro-sh}\}$. Interpretation is aided by the ability to visualize geodesic paths between cell nuclei, whose lengths correspond to the computed distances.

\subsection{Second-order analysis and practical implications}
\begin{figure}[!t]
    \centering
    \begin{tabular}{@{}c@{}c@{}c@{}}
             \includegraphics[width=0.33\linewidth]{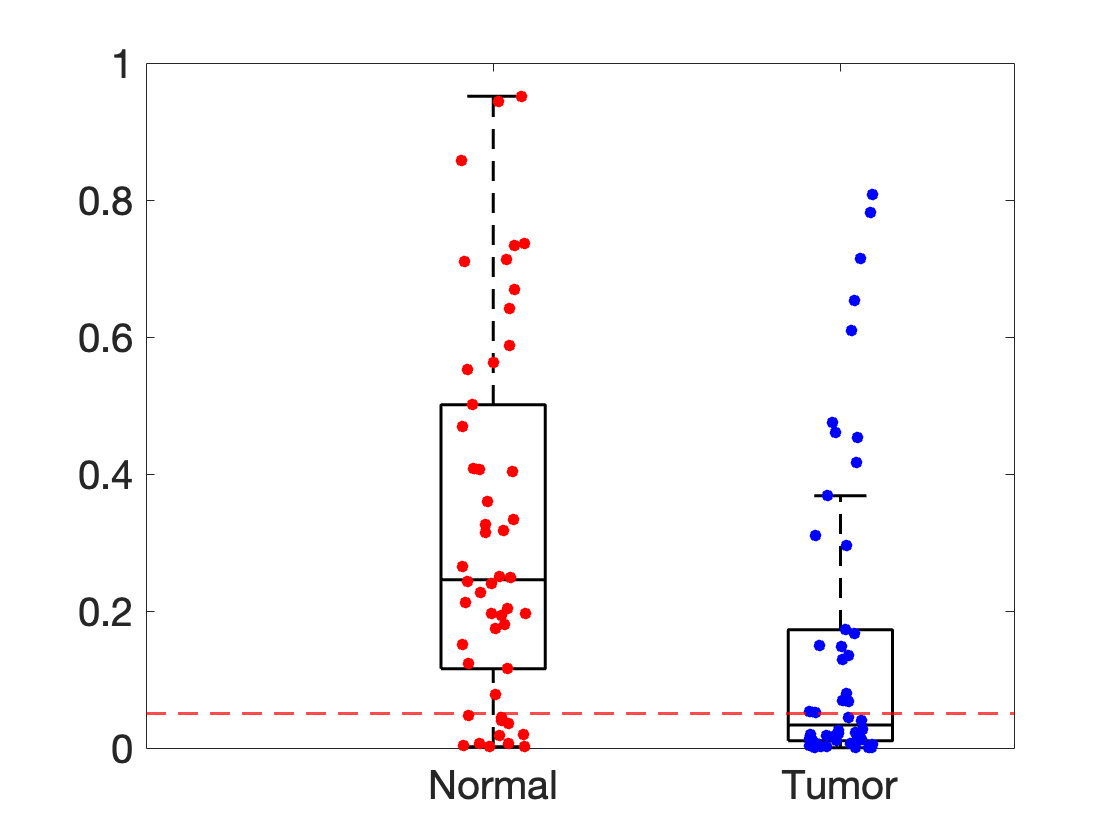} & \includegraphics[width=0.33\linewidth]{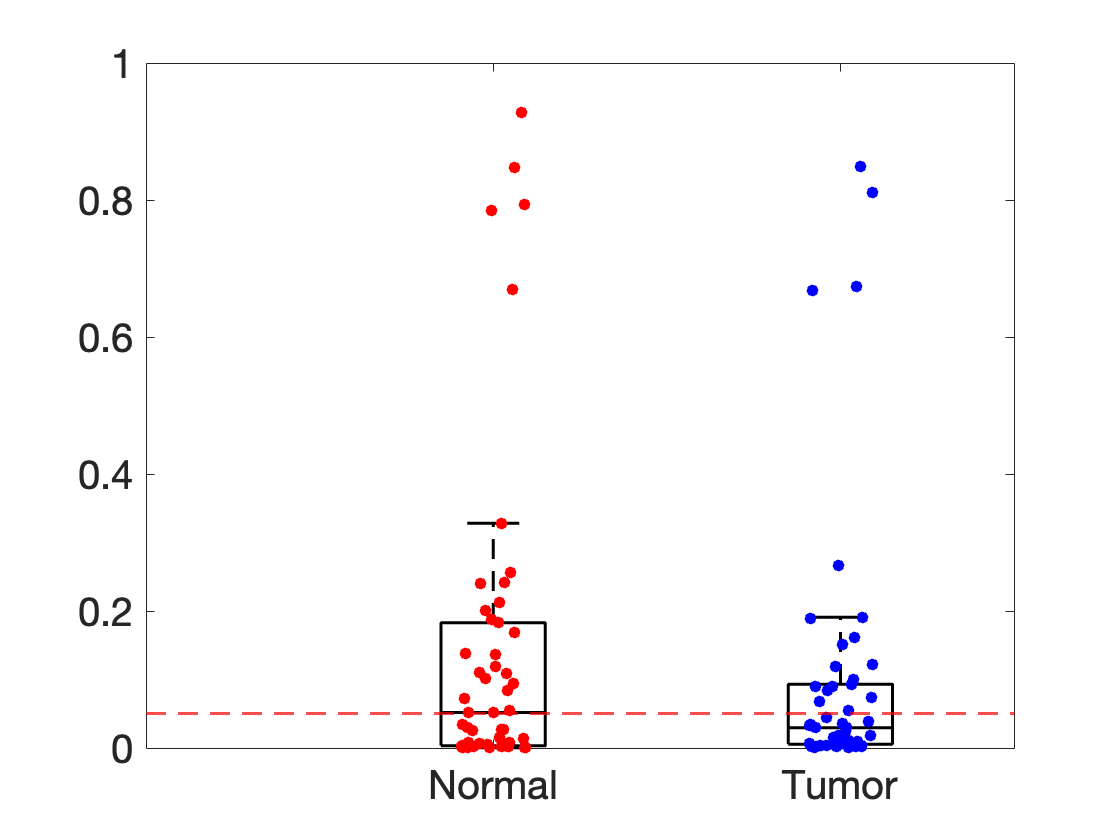} &
         \includegraphics[width=0.33\linewidth]{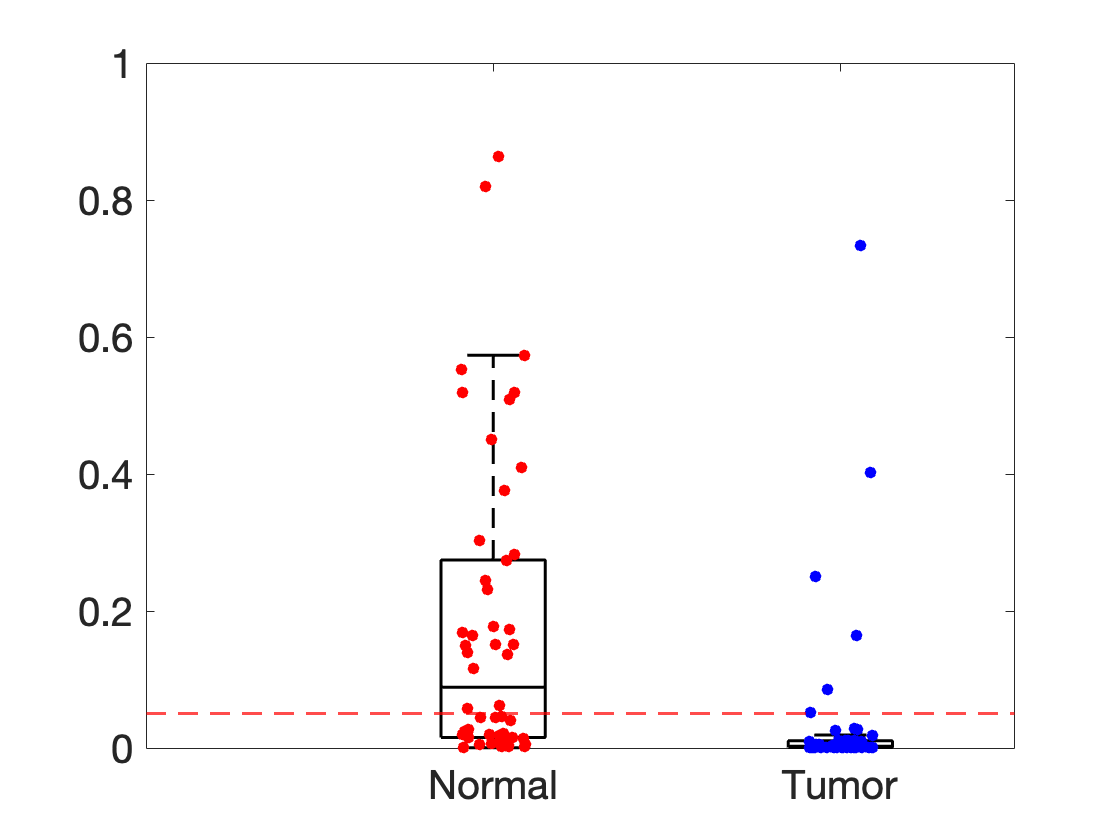} \\
         \hline
         \includegraphics[width=0.33\linewidth]{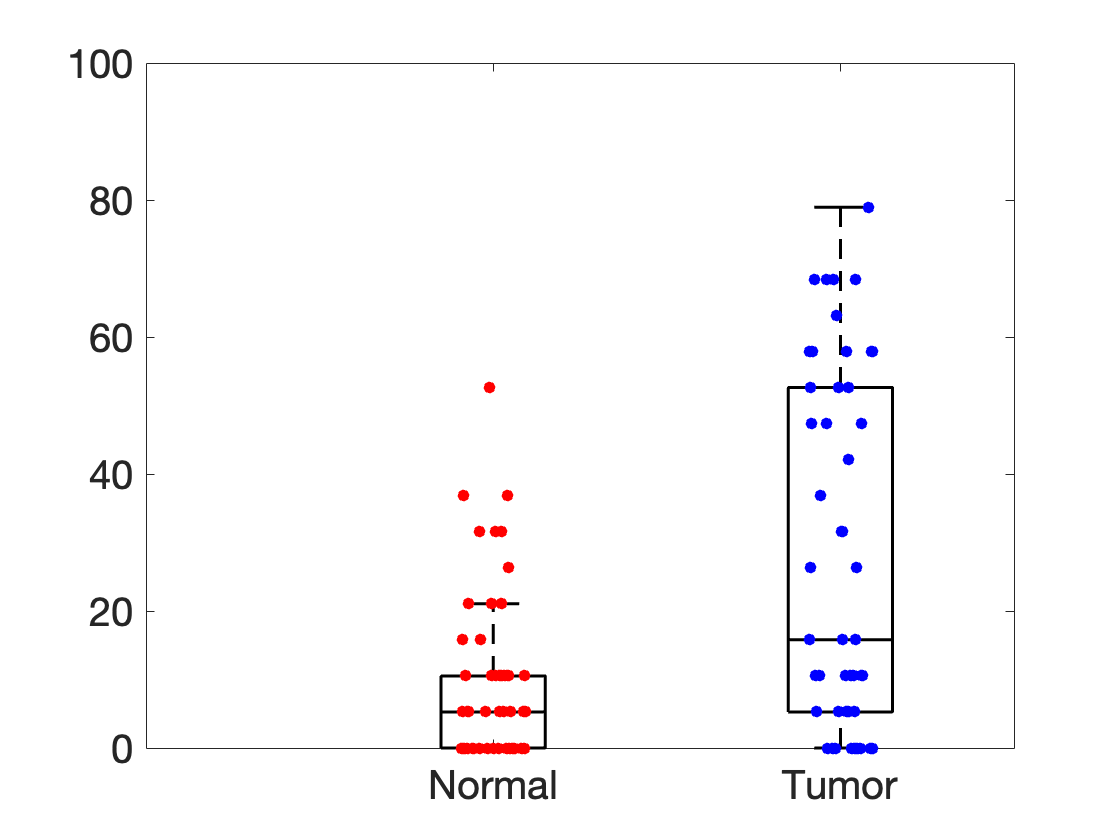} & \includegraphics[width=0.33\linewidth]{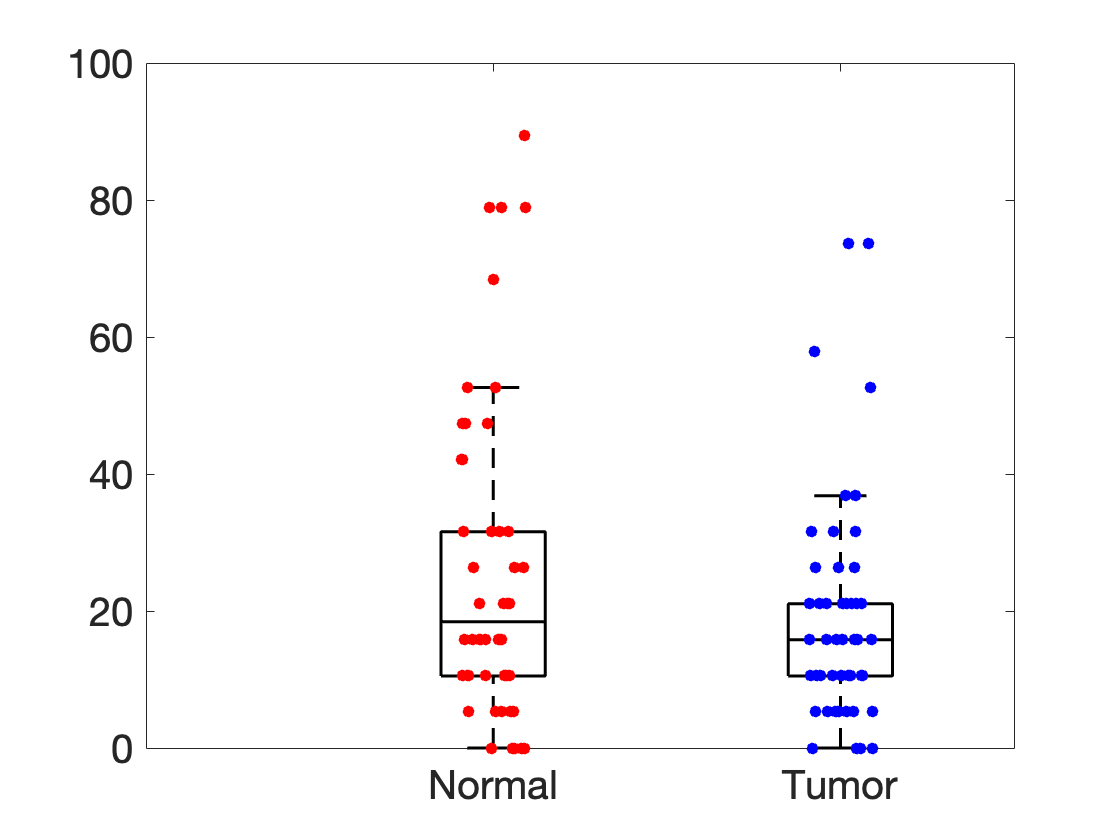} &
         \includegraphics[width=0.33\linewidth]{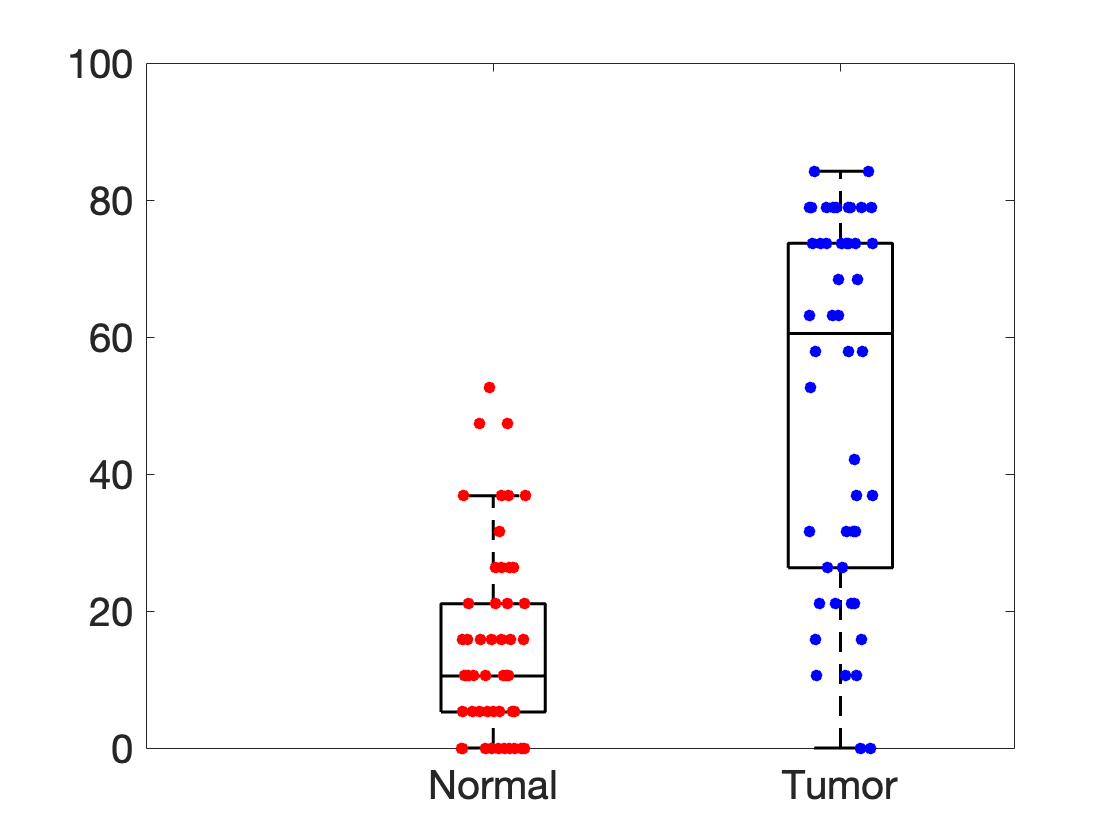} \\
         (a)&(b)&(c)
    \end{tabular}
    \caption{\small Boxplots of p-values from the global rank envelope test (top), and proportions of radii for which the mark-weighted $K$ functions deviated outside the 95\% pointwise envelope (bottom), based on the nuclei (a) shapes, (b) orientation-and-shapes, and (c) size-and-shapes. The red dashed line represents the significance level $\alpha = 0.05$ used in the global test. Points within each boxplot were jittered along the $x$-axis for clarity.}
    \label{fig:boxplot-breast-cancer}
\end{figure}

For each sampled region, we first estimated mark-weighted $K$ functions, based on shape, orientation-and-shape or size-and-shape test functions. The mark-weighted $K$ functions were then used to carry out (i) global rank envelope tests (using extreme rank depth) at a significance level of $\alpha=0.05$, and (ii) local tests using 95\% pointwise envelopes. P-values from the global tests are shown in the first row of Figure \ref{fig:boxplot-breast-cancer}. The second row reports the proportions (converted to \%) of radii $r$ for which the $K$ functions deviated from the pointwise envelope. In all cases, we also show the boxplots of the p-values or proportions. Panels (a)-(c) correspond to results for shape, orientation-and-shape and size-and-shape, respectively. Interesting patterns immediately emerge when examining these plots. In (a) and (c), it is evident that there is strong evidence of spatial dependence among the nuclei shapes and size-and-shapes for many more regions within tumor tissue as compared to non-tumor tissue. For normal tissue regions, the shape mark-weighted $K$ function deviated form the pointwise envelope for small proportions radii: most between 0\% and 20\%. On the other hand, for tumor tissue regions, the proportions are much higher. This is further supported by the p-values from the global tests, which are much lower overall for tumor tissue regions (often lower than $\alpha=0.05$). This pattern is even stronger when the size-and-shape mark-weighted $K$ function is employed. This indicates that cell nuclei shapes and size-and-shapes are much more correlated with their spatial locations within tumor tissue than within non-tumor tissue. It further appears that cell nuclei within tumor and non-tumor tissues exhibit similar (strength of) spatial correlation in terms of orientation-and-shape. 

\begin{figure}[!t]
    \centering
    \begin{tabular}{@{}c@{}c@{}c@{}}
         \multicolumn{3}{c}{\includegraphics[width=0.3\linewidth]{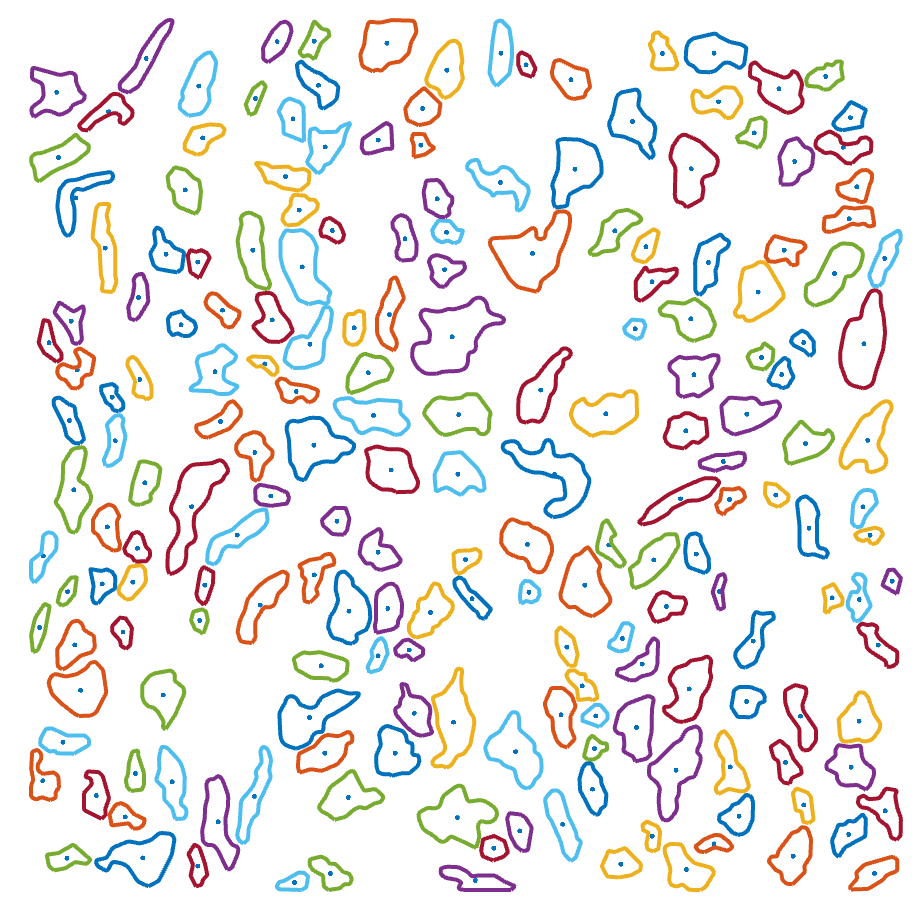 }} \\ 
         \multicolumn{3}{c}{(a)}\\
         \includegraphics[width=0.33\linewidth]{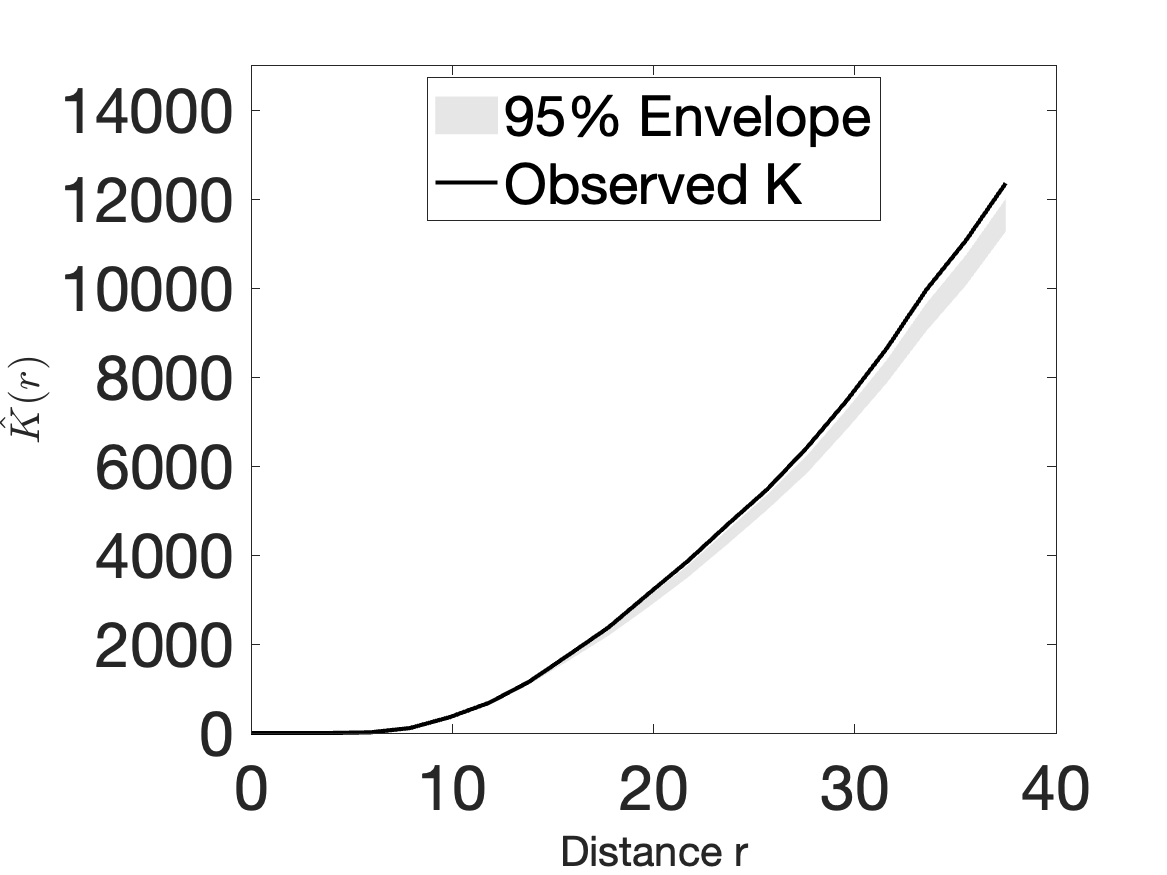} &
         \includegraphics[width=0.33\linewidth]{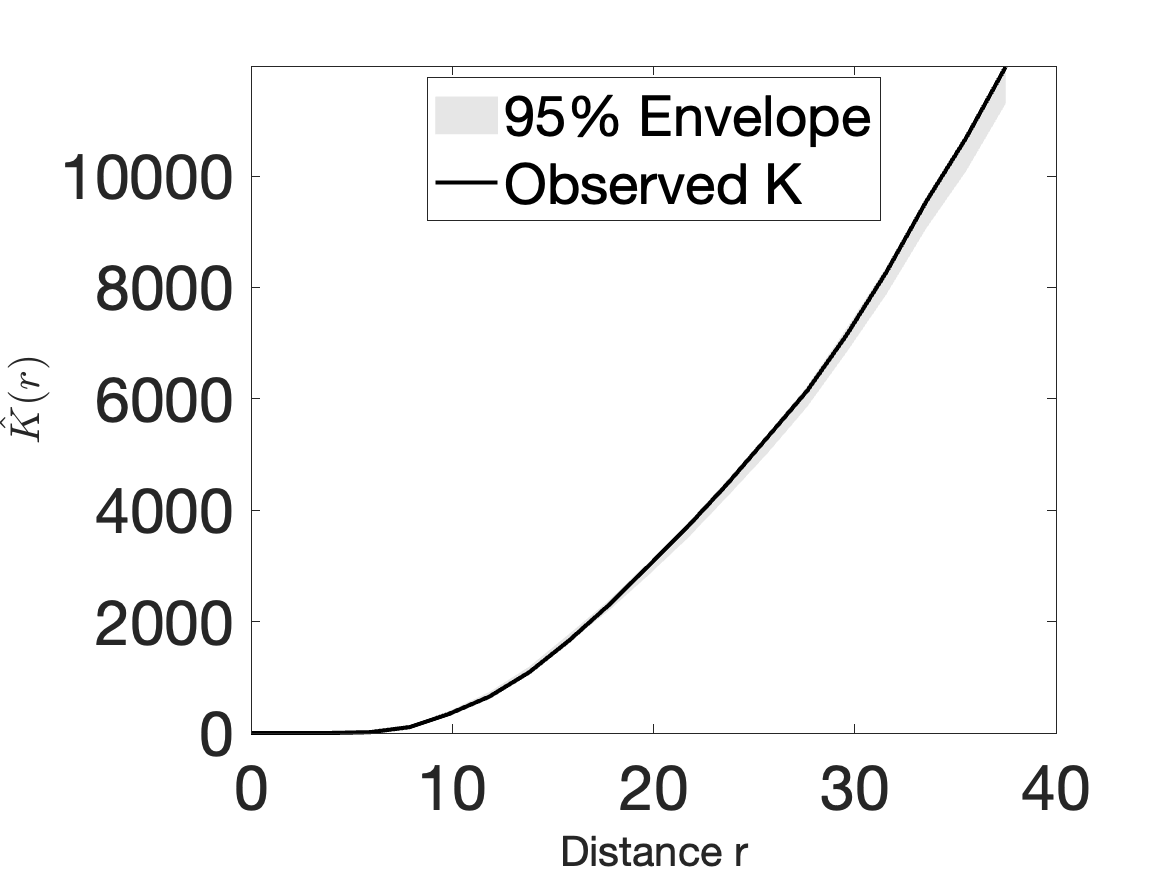} &
         \includegraphics[width=0.33\linewidth]{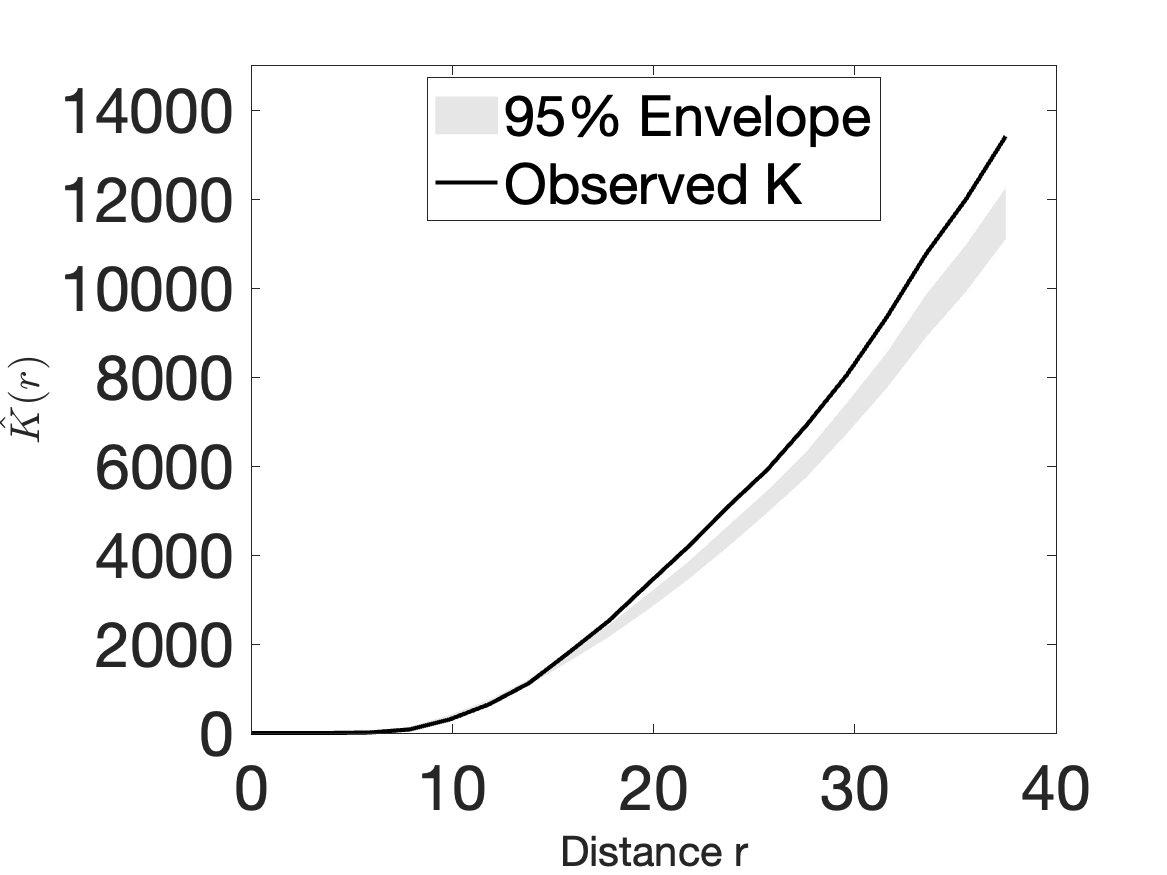}\\
         (b)&(c)&(d)\\
         \includegraphics[width=0.33\linewidth]{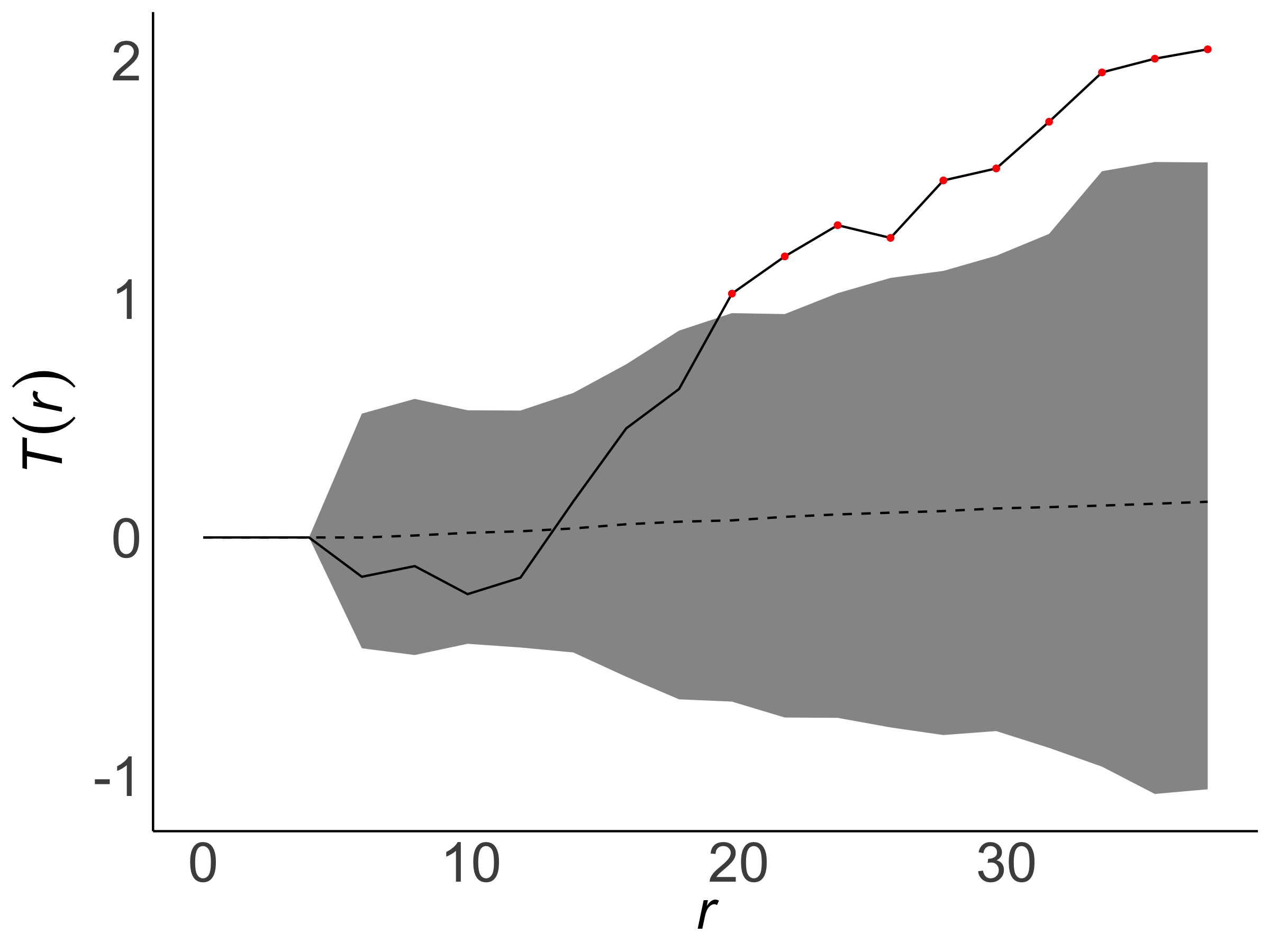} & \includegraphics[width=0.33\linewidth]{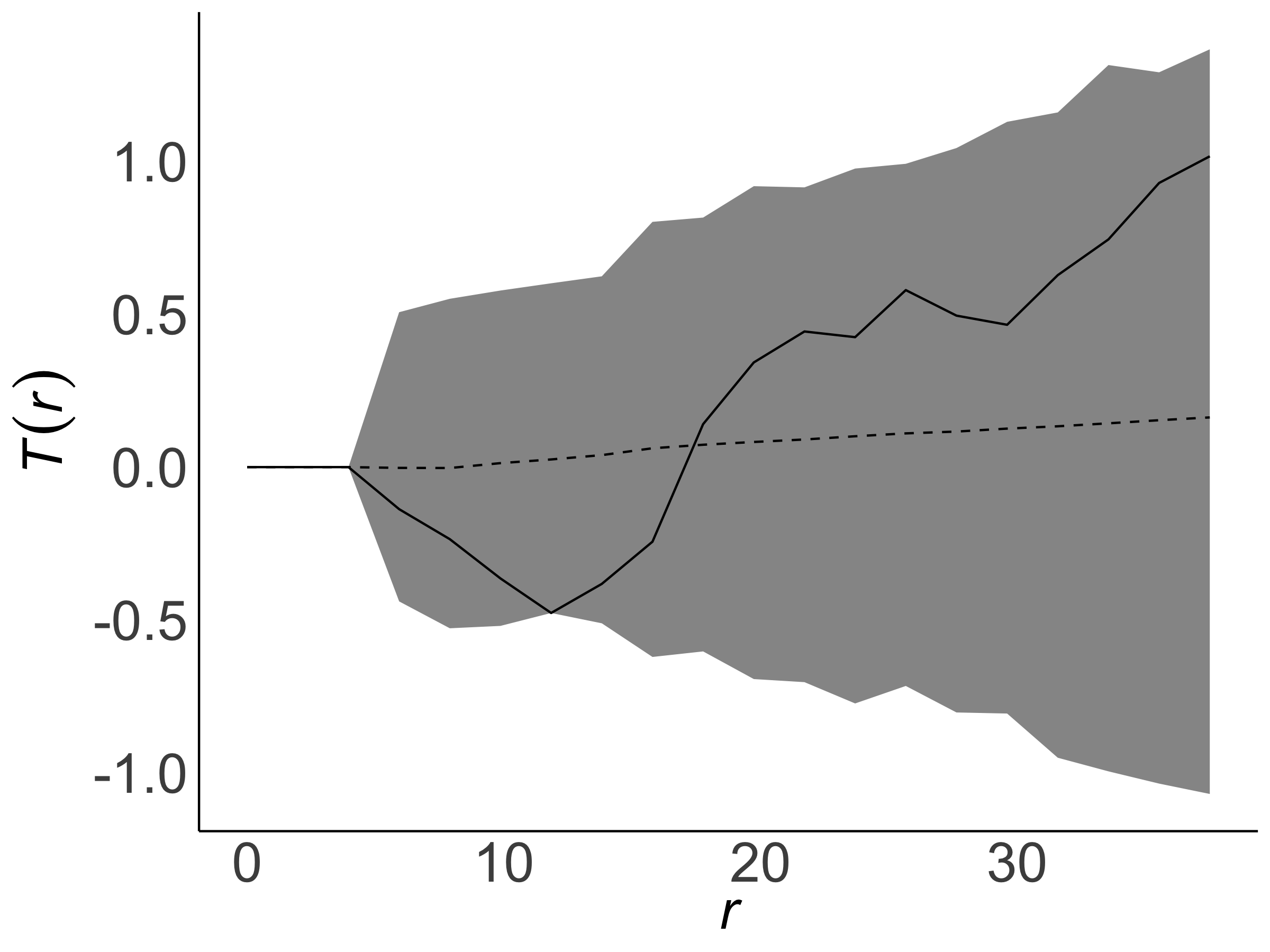} &
         \includegraphics[width=0.33\linewidth]{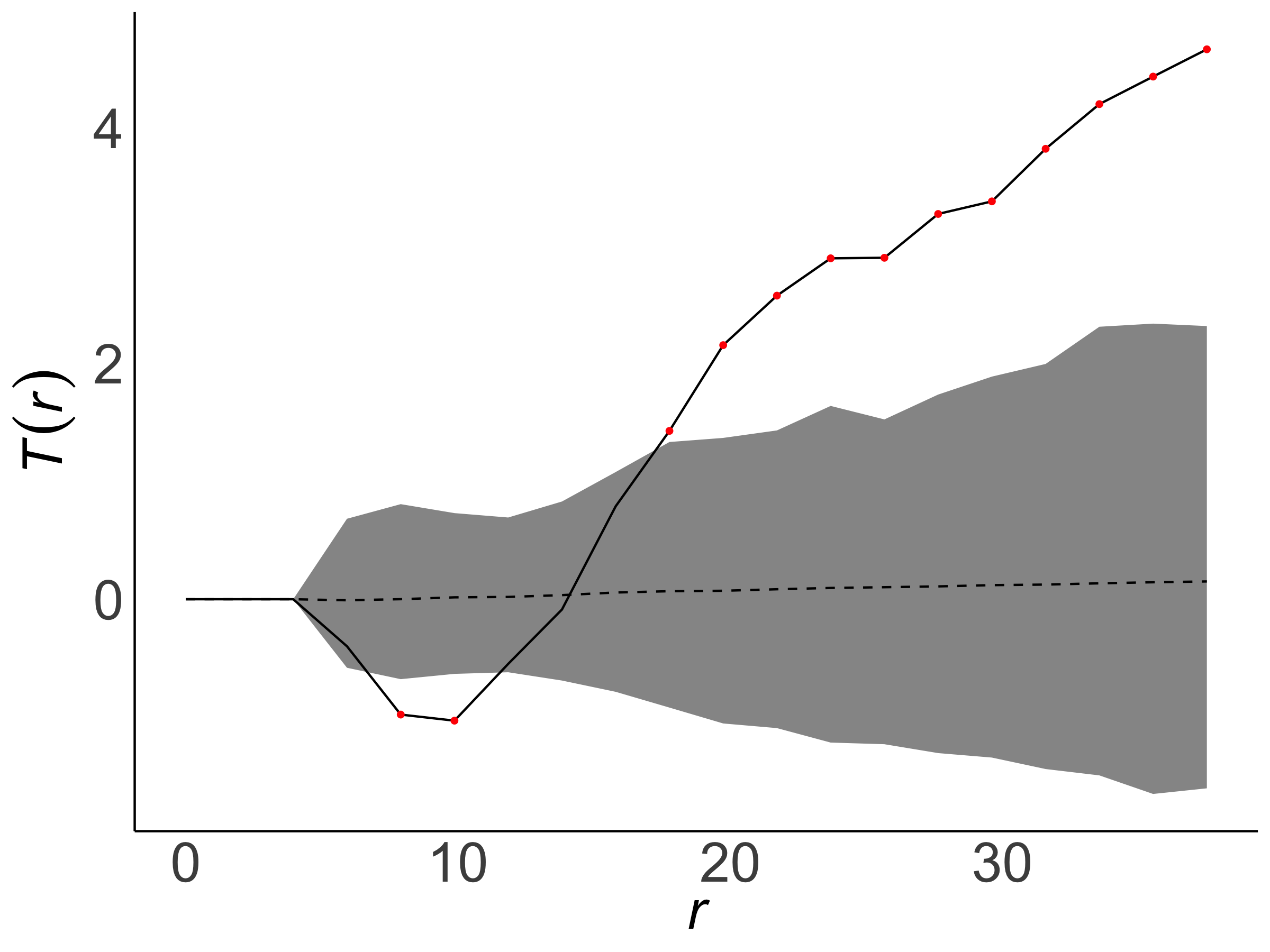}\\
         (e)&(f)&(g)
    \end{tabular}
   \caption{\small (a) A region within tumor tissue. Mark-weighted $K$ and $L$ functions along with their pointwise and global envelopes, respectively, estimated using (b)\&(e) shape, (c)\&(f) orientation-and-shape, and (d)\&(g) size-and-shape test functions.}
    \label{fig:k-breast-cancer-tumor}
\end{figure}

Figure \ref{fig:k-breast-cancer-tumor} shows how spatial dependence based on various types of variation emerges in a randomly chosen region within tumor tissue. Panel (a) presents the nuclei outlines. Panels (b)-(d) show the shape, orientation-and-shape and size-and-shape mark-weighted $K$ functions, respectively, along with their 95\% pointwise envelopes. Panels (e)-(g) display the corresponding test statistics used for the global tests along with their 95\% global envelopes. In panel (e), it is evident that nuclei shapes do not exhibit spatial correlation up to approximately $r=20$, after which they become more variable than expected under random labeling (deviation above the global envelope). In panel (g), the nuclei size-and-shapes exhibit smaller variability than expected by random labeling for radii close to $r=10$. However, after approximately $r=20$, they are more variable than expected under random labeling. In panel (f), we do not see evidence against random labeling based on the nuclei orientation-and-shapes. We observed very similar spatial dependence patterns in the other regions that were sampled within tumor tissues. For regions sampled in non-tumor tissues, the mark-weigthed $K$ functions (shape, orientation-and-shape or size-and-shape) did not consistently deviate from the pointwise or global envelopes; see Section 6 in the supplement. 

The spatial dependency patterns for shape and size-and-shape aligned with expected tumor cell behavior. The mark-weighted $K$ functions consistently remained within the expected range (envelope) for non-tumor cell nuclei. On the other hand, particularly based on size-and-shape, tumor cell nuclei showed significant deviations: values dropped below the envelope for radii of 6-13 $\mu m$, then rose above the envelope starting at a median radius of approximately 15.5 $\mu m$ (the median radius was computed across all regions sampled within tumor tissues). Based on shape alone, the deviations above the envelope started at a median radius of approximately 17.5 $\mu m$. For reference, patient-derived circulating breast cancer cells have a median diameter of 12.4 $\mu m$ \citep{Mendelaar2020DefiningThe}; note that the cell nuclei are smaller in size. The deviation of the size-and-shape mark-weighted $K$ functions below the envelope for small radii likely reflects tumor cells' distinct spatial organization. Normal breast cells arrange in tubular formations around ducts, whereas tumor cells form tight clusters of genetically similar `clones' \citep{GreavesMel2012Ceic}. This clustering of morphologically similar cells at short distances produces the observed reduction in $K$ function values, as neighboring cells exhibit less variation than would be expected by random labeling.

For larger radii, where clustering effects diminish, the significantly elevated $K$ function values might reflect tumor heterogeneity -- the diverse shapes and/or sizes of cancer cells within the same tumor. Unlike normal cells that maintain consistent nuclear morphology due to genetic stability, tumor cells accumulate different mutations and experience varying local conditions, creating a mixed population with markedly different appearances. This morphological diversity among neighboring cells results in more variation than expected under random labeling \citep{Alizadeh2015,zardavas2015clinical,MarusykAndriy2012Ihal}.

\section{Concluding remarks}
\label{sec:discussion}
Motivated by the problem of uncovering and testing for spatial dependence patterns among cell nuclei shapes in histopathology images of cancer, we propose a marked point process framework based on a mark-weighted $K$ function statistic to capture shape correlations among cell nuclei. The framework uses the elastic shape analysis of curves and ensures invariance to desired symmetries in the underlying data. Simulations confirm the effectiveness of the novel $K$ functions in capturing relevant spatial correlations for planar closed curves, and application of the proposed framework to histopathology images of breast cancer reveals distinct spatial dependence patterns in their shape and size-and-shape across tumor and non-tumor regions. Our findings corroborate existing hypotheses in the literature on breast cancer tumor growth and heterogeneity, as well as cell proliferation.

We have identified several directions for future work. First, we will build on the framework of \cite{ghorbani2021functional} to quantify spatial relationships among shape marks modulated by auxiliary variables in a latent space. Second, we will extend the framework to apply to shapes of outlines of three-dimensional objects, i.e., surfaces. Cell nuclei are inherently three-dimensional and this will allow us to capture full information about their shape. Third, we will try to establish consistency of the proposed mark-weighted $K$ functions when the reparameterization group is restricted to a finite-dimensional family; see discussion in Section 4 in the supplement. Finally, we plan to apply the proposed methods to other types of cancer, e.g., prostate, to try to understand what properties of cell nuclei affect their spatial distribution. This has the potential to reveal different types of tumor heterogeneity in different contexts.\\

\if1\blind
{\noindent\textbf{Acknowledgments.} This work was supported in part by NIH R37-CA214955 (SK and KB), NSF DMS-2015374 and EPSRC EP/Z003377/1 (KB), and NSF DMS-2413747 (SK).
} \fi

\if0\blind
{
} \fi

\bibliographystyle{plainnat}
\bibliography{ref}
\end{document}


\def\spacingset#1{\renewcommand{\baselinestretch}%
{#1}\small\normalsize} \spacingset{1}


\if1\blind
{
  \title{\bf Supplement for second-order spatial analysis of shapes of tumor cell nuclei}
  \author{Ye Jin Choi, Sebastian Kurtek\\
    Department of Statistics, The Ohio State University\\
    Simeng Zhu\\
    James Cancer Center, The Ohio State University\\
    Karthik Bharath \\
    School of Mathematical Sciences, University of Nottingham}
  \maketitle
} \fi

\if0\blind
{
  \bigskip
  \bigskip
  \bigskip
  \begin{center}
    {\LARGE\bf Supplement for second-order spatial analysis of shapes of tumor cell nuclei}
\end{center}
  \medskip
} \fi

\spacingset{1.55} 

\section{Reference measure on the mark space of shapes}
The space $\mathcal Q$ is a codimension two submanifold of $\mathbb L^2(\mathbb S^1, \mathbb R^2)$, and a Polish space. Let $\zeta_1$ be the restriction of a non-degenerate Gaussian measure on $\mathbb L^2(\mathbb S^1, \mathbb R^2)$ to $\mathcal Q$. Consider first only the reparametrization group $\Gamma$ as the symmetry. 
The reparametrization group $\Gamma$ may be prescribed an appropriate topology under which it is a Polish space \citep{cohen2017polishability}. Let $\zeta_2$ be a quasi-invariant $\sigma$-finite measure on $\Gamma$ \citep[see, e.g., Section 11.5][]{bogachev2010differentiable}. The group action $\Phi: \mathcal Q \times \Gamma \to \mathcal Q,\ (q,\gamma) \to (q\circ \gamma) \sqrt{\dot \gamma}$ is continuous. Thus, the pushforward $\nu:=\zeta_1 \times \zeta_2 \circ \Phi^{-1}$ of the product measure, also known as the convolution measure, on $\mathcal Q \times \Gamma$ is well-defined, with support on a closed subset of $\mathcal Q$. 

The rotation group $SO(2)$ is a compact Lie group with a single connected component, and Polish. Let $\zeta_3$ be the uniform Haar measure on $SO(2)$. The action $(Oq,\gamma) \to  O(q\circ \gamma) \sqrt{\dot \gamma}$ is also continuous, which ensures that the pushforward of the product measure $\zeta_1 \times \zeta_2 \times \zeta_3$ under the group action results similarly in a well-defined measure with support on a closed subset of $\mathcal Q$. 

\section{Normalizing constant in population $K$ function}

We show that, under random labeling, the normalizing constant takes the form $c_\ff = \mathbb{E}\left( \left\| q * g - m_G \right\|^2 \right)$, where $m_G = \mathbb{E}(q * g)$ and $g \in G$.

\begin{align*}
    c_\ff &:= \mathbb{E} \left[ \ff(q * g, q_1 * g_1) \mid (q * g, q_1 * g_1) \perp X \right] \\
    &= \mathbb{E}\left( \frac{1}{2} \left\| q * g - q_1 * g_1 \right\|^2 \right) \\
    &= \frac{1}{2} \, \mathbb{E} \left( \left\| (q * g - m_G) + (m_G - q_1 * g_1) \right\|^2 \right) \\
    &= \mathbb{E} \left( \left\| q * g - m_G \right\|^2 \right) 
    + \int_{\mathbb{S}^1} \mathbb{E} \left[ \left( (q * g)(t) - m_G(t) \right) \left( (q_1 * g_1)(t) - m_G(t) \right) \right] \text{d}t \\
    &= \mathbb{E} \left( \left\| q * g - m_G \right\|^2 \right)
\end{align*} The fourth equality follows from Fubini’s theorem with $\text{d}t$ as the arc length measure. The final equality holds because the marks \( q * g \) and \( q_1 * g_1 \) are independent under the assumption of random labeling.

\section{Unbiasedness of the estimator $\bar K_\ff$}
The estimator $\bar K_\ff$ in Equation (6) in the main manuscript is unbiased when the symmetries $g$ are \emph{known} or are exactly recovered. To see this, consider first the following intermediate result, which is a special case of Lemma 1 in \cite{ghorbani2021functional} with an additional normalizing factor $c_\ff$. Let $\kappa_\ff (u_1,u_2)$ denote the 2nd order intensity-reweighted $t$-correlation measure (Definition 8 in \cite{ghorbani2021functional}), defined by
\[
\kappa_\ff(u_1,u_2) = \int_{\mathcal{M}^2} \frac{J_{u_1,u_2}}{J_{u_1}J_{u_2}} \, \ff(q_1*g_1,q_2*g_2) \, \prod_{i=1}^2 \dd\nu(q_i*g_i).
\] Then, the $K$ function can be expressed as
\begin{equation}
K_\ff (r) = \frac{1}{c_{\ff}}\int_{B_0(r)} \kappa_\ff(0,u_1) \, K_\text{gr}(\dd u_1),
\label{eq:inter-result}    
\end{equation}
where $K_\text{gr}$ is the 2nd order reduced moment measure of the ground process.

Let $\mathbb E^{!(x,(q,\gamma))}$ denote the expectation under the reduced Palm measure $P^{!{(x,(q,\gamma))}}$, which is the conditional distribution of a functional marked point process given that the it has a point at $(x,(q*g))$. Then, we can express $c_\ff K_\ff(r)$ as
\begin{align*}
&c_\ff K_\ff (r)=\frac{1}{|\mathcal X|}\mathbb{E}\left[ \sum_{(x,(q*g))\in\Psi}\sum_{(x_1,(q_1*g_1) \in \Psi \setminus \{(x,(q*g))\}} \frac{\ff(q*g,q_1*g_1)}{\rho(x, q*g)} \frac{\mathbb I\{x_1\in \mathcal X \cap B_{x}(r)\}}{\rho(x_1,q_1 *g_1)}\right]\\
&=\frac{1}{|\mathcal X|}\int_{\mathcal{X}\times \mathcal M}\mathbb E^{!(x,q*g)} \left[\sum_{(x_1,q_1*g_1) \in \Psi \setminus \{(x, q*g)\}} \frac{\ff(q*g,q_1*g_1)}{\rho(x, q*g)} \frac{\mathbb I\{x_1\in \mathcal X \cap B_{x}(r)\}}{\rho(x_1,q_1 *g_1)} \right] \dd \lambda(x ,q*g).
\end{align*}
The last equality is due to the Campbell-Mecke formula. Further, $\mathbb E^{!(x,q*g)}(\cdot)$ can be expressed as
\begin{align*}
&\mathbb{E}^{!(x,q*g)} \left[\sum_{(x_1,q_1*g_1) \in \Psi \setminus \{(x, q*g)\}} \frac{\ff(q*g,q_1*g_1)}{\rho(x, q*g)} \frac{\mathbb I\{x_1\in \mathcal X \cap B_{x}(r)\}}{\rho(x_1,q_1 *g_1)} \right]\\
&=\int_{\mathcal X \times \mathcal M}\frac{\ff(q*g,q_1*g_1)}{\rho(x, q*g)} \frac{\mathbb I\{x_1\in \mathcal X \cap B_{x}(r)\}}{\rho(x_1,q_1 *g_1)}\rho^{(2)}((x,q*g),(x_1,q_1*g_1)) \dd \lambda (x_1,q_1*g_1)\\
&=\int_{\mathcal X \times \mathcal M}\ff(q*g,q_1*g_1) \mathbb I\{x_1 \in \mathcal X \cap B_{x}(r)\} \eta_{\Psi}((x,q*g),(x_1,q_1*g_1))\dd \lambda (x_1,q_1*g_1)\\
&\overset{\text{A1}}{=}\int_{\mathcal X \times \mathcal M}\ff(q*g,q_1*g_1) \mathbb I\{x_1 \in \mathcal X \cap B_{x}(r)\} \eta_{\Psi}((0,q*g),(x_1-x,q_1*g_1))\dd \lambda (x_1,q_1*g_1)\\
&\overset{u_1=x_1-x}{=} \int_{B_{0}(r)\times\mathcal{M}}\ff(q*g, q_1*g_1)\eta_{\Psi}((0,q*g),(u_1,q_1*g_1))\dd \lambda (u_1,q_1*g_1)\\
&=\int_{B_0(r)\times\mathcal{M}}\ff(q*g, q_1*g_1)\frac{J_{0,u_1}(q*g,q_1*g_1)}{J_0(q*g) J_{u_1}(q_1*g_1)} \eta_{\text{gr}}(0,u_1)\dd \lambda (u_1,q_1*g_1)\\
&=\int_{B_0(r)}\Bigg( \int_{\mathcal{M}}\ff(q*g, q_1*g_1)\frac{J_{0,u_1}(q*g,q_1*g_1)}{J_0(q*g) J_{u_1}(q_1*g_1)} \dd \nu(q*g)\Bigg)\eta_{\text{gr}}(0,u_1)\dd u (u_1).
\end{align*}
Thus, we have
\begin{align*}
&c_\ff K_\ff (r)\\
&=\frac{1}{|\mathcal X|}\int_{\mathcal X\times\mathcal{M}}\mathbb E^{!(x,q*g)} \left[\sum_{(x_1,q_1*g_1) \in \Psi \setminus \{(x, q*g)\}} \frac{\ff(q*g,q_1*g_1)}{\rho(x, q*g)} \frac{\mathbb I\{x_1 \in \mathcal X \cap B_x(r)\}}{\rho(x_1,q_1*g_1)} \right]\dd\lambda(x,q*g)\\
&=\frac{1}{|\mathcal X |}\int_{\mathcal X \times \mathcal M}\int_{B_0(r)}\Bigg( \int_{\mathcal{M}}\ff(q*g, q_1*g_1)\frac{J_{0,u_1}(q*g,q_1*g_1)}{J_0(q*g) J_{u_1}(q_1*g_1)} \dd \nu(q*g)\Bigg)\times\\
& \qquad \times \eta_{\text{gr}}(0,u_1)\dd u (u_1)\dd\lambda(u,q*g)\\
&= \frac{1}{|\mathcal X |}\int_{\mathcal X}\int_{B_0(r)}\Bigg( \int_{\mathcal M^2}\ff(q*g, q_1*g_1)\frac{J_{0,u_1}(q*g,q_1*g_1)}{J_0(q*g) J_{u_1}(q_1*g_1)} \dd \nu(q*g) \dd \nu(q_1*g_1) \Bigg)\times \\
& \qquad \times \eta_{\text{gr}}(0,u_1)\dd u (u_1)\dd u(u)\\
&= \frac{1}{|\mathcal X |}\int_{\mathcal X}\int_{B_0(r)}\kappa_\ff (0,u_1) \eta_{\text{gr}}(0,u_1)\dd u (u_1)\dd u(u)\\
&= \frac{1}{|\mathcal X |}\int_{\mathcal X} \dd u(u) \int_{B_0(r)}\kappa_\ff (0,u_1) \eta_{\text{gr}}(0,u_1)\dd u (u_1)\\
&=\int_{B_0(r)}\kappa_\ff(0,u_1) K_\text{gr}(\dd u_1).
\end{align*}
Then,
\begin{align*}
    &\mathbb{E}\left[c_\ff\bar K_\ff(r)\right]\\
    &=\frac{1}{|\mathcal X|} \mathbb{E}\Bigg[ \sum_{(x,q*g)\in\Psi}\sum_{(x_1,q_1*g_1)\in\Psi\setminus\{(x,q*g)\}} w(x,x_1)\ff(q*g, q_1*g_1) \times\\
    &\qquad \times \frac{\mathbb I\{x\in \mathcal X\}}{\rho(x,q*g)}\frac{1\{x_1\in \mathcal X\cap B_x(r))\}}{\rho(x_1,q_1*g_1)}\Bigg]\\
    &=\frac{1}{|\mathcal X|}\int_{\mathcal{M}\times \mathcal X}\int_{\mathcal{M}\times \mathcal X} w(x,x_1)\ff(q*g, q_1*g_1)
    \frac{\mathbb I\{x\in \mathcal X\}}{\rho(x,q*g)}\frac{1\{x_1\in \mathcal X \cap B_x(r)\}}{\rho(x_1,q_1*g_1)} \times \\ 
    & \qquad \times \rho^{(2)}((x,q*g),(x_1,q_1*g_1))
    \dd \lambda (x,q*g)\dd \lambda (x_1,q_1*g_1)
\end{align*}
\begin{align*}
    &=\frac{1}{|\mathcal X|} \int_{\mathcal{M}}\int_{\mathcal{M}} \ff(q*g,q_1*g_1) \times \\
    & \qquad \times \left(\int_\mathcal{X} \int_{\mathcal X \cap B_x(r)}w(x,x_1)\frac{\rho^{(2)}((x,q*g),(x_1,q_1*g_1))}{\rho(x,q*g)\rho(x_1,q_1*g_1)} \dd x(x_1) \dd x(x) \right) \times \\
    & \qquad \times 
    \dd \nu (q*g) \dd \nu(q_1*g_1)\\
    &=\frac{1}{|\mathcal X|}\int_{\mathcal{M}}\int_{\mathcal{M}}  \ff(q*g,q_1*g_1) \times \\
    & \qquad \times \left(\int_\mathcal{X} \int_{\mathcal X \cap B_x(r)}w(x,x_1)\eta_{\Psi}((x,q*g),(x_1,q_1*g_1)) \dd x(x_1) \dd x(x) \right) \times \\
    &\qquad \times \dd \nu (q*g) \dd \nu(q_1*g_1),
\end{align*}
and by assumption A1 and Fubini's theorem, the inner expression satisfies 
\begin{align*}
    &\int_\mathcal{X} \int_{\mathcal X \cap B_x(r)}w(x,x_1)\eta_{\Psi}((x,q*g),(x_1,q_1*g_1)) \dd x(x_1) \dd x (x)\\
    &\overset{\text{A1}}{=}\int_\mathcal{X} \int_{\mathcal X}\mathbb I \{x_1\in \mathcal X \cap B_x(r)\}w(x,x_1)\eta_{\Psi}((0,q*g),(x_1-x,q_1*g_1)) \dd x(x_1) \dd x(x)\\
    &\overset{u_1=x_1-x}{=}\int_{\mathcal X}\int_\mathcal{X} \mathbb I \{u_1+x\in \mathcal X \cap B_x(r)\}w(x,x+u_1)\dd x(x) \times \\
    & \qquad \times \eta_{\Psi}((0,q*g),(u_1,q_1*g_1))\dd u(u_1)\\
    &=\int_{\mathcal X}\mathbb I \{u_1\in B_0(r)\}\left(\int_\mathcal{X} \mathbb I \{u_1+x\in \mathcal X \}w(x,x+u_1)\dd x(x)\right) \times \\
    & \qquad \times \eta_{\Psi}((0,q*g),(u_1,q_1*g_1))\dd u(u_1)\\
     &=|\mathcal X|\int_{B_0(r)} \eta_{\Psi}((0,q*g),(u_1,q_1*g_1))\dd u(u_1),
\end{align*}
because
$$
\int_\mathcal X \mathbb I\{u_1+x\in \mathcal X\}w(x,x+u_1)\dd x (x) = |\mathcal X |.
$$
Hence, by Fubini's theorem and the intermediate result in \eqref{eq:inter-result}, we have the following:
\begin{align*}
&\frac{1}{|\mathcal X|}\int_{\mathcal{M}}\int_{\mathcal{M}} \ff(q*g,q_1*g_1) \left(\int_\mathcal X \int_{\mathcal X \cap B_x(r)} w(x,x_1)\eta_{\Psi}((x,q*g),q_1*g_1)) \dd x(x_1) \dd x (x) \right)\\
& \qquad \times \dd \nu(q*g) \dd \nu(q_1*g_1)\\
&=\frac{1}{|\mathcal X|}\int_{\mathcal{M}}\int_{\mathcal{M}}\ff (q*g,q_1*g_1)\left(|\mathcal X|\int_{B_0(r)}\eta_{\Psi}((0,q*g),(u_1,q_1*g_1))\dd u(u_1)\right)\\
& \qquad \times \dd \nu(q*g)\dd \nu(q_1*g_1)\\
&= \int_{\mathcal{M}}\int_{\mathcal{M}}\ff(q*g,q_1*g_1) \left(\int_{B_0(r)} \frac{J_{0,u_1}(q*g,q_1*g_1)}{J_0(q*g) J_{u_1}(q_1*g_1)} \eta_\text{gr}(0,u_1)\dd u(u_1)\right)\\
& \qquad \times \dd \nu(q*g)\dd \nu(q_1*g_1)\\
&\overset{Fubini}{=}\int_{B_0(r)}\left[ \int_{\mathcal{M}}\int_{\mathcal{M}} \frac{J_{0,u_1}(q*g,q_1*g_1)}{J_0(q*g) J_{u_1}(q_1*g_1)} \ff(q*g,q_1*g_1)\dd \nu (q*g)\dd \nu(q_1*g_1))\right] \\
& \qquad \times \eta_\text{gr}(0,u_1) \dd u(u_1)\\
&=\int_{B_0(r)} \kappa_\ff(0,u_1)\eta_\text{gr} (0,u_1)\dd u(u_1)\\
&=\int_{B_0(r)} \kappa_\ff(0,u_1)K_\text{gr}(\dd u_1)\\
&\overset{\eqref{eq:inter-result}}{=}c_\ff K_\ff(B_0(r))\\
&\equiv c_\ff K_\ff(r)
\end{align*}

\section{Challenges in establishing consistency of $\hat K$}
\label{sec:consistency}

Pointwise consistency of the estimator $\hat K$ depends on, for each location $x \in \mathcal X$, the exact recovery of the symmetries $g, g_1$, and values of the ground intensity $\rho_{\text{gr}}$ at $x$ and $x_1$; then, continuity of the distance functions $d_j,\ j \in \{\text{sh},\text{sc-sh},\text{ro-sh}\}$ ensures convergence of $\hat \ff$ to $\ff$, and from the law of large numbers in Hilbert spaces (since an extrinsic metric $\|\cdot\|$ is used on $\mathcal Q$), $\hat c_\ff$ also converges to $c_\ff$ under random labeling. 

However, exact recovery of symmetries is not possible in practice. The key issue lies with the infinite-dimensionality of the symmetry group due to the presence of the reparametrization group $\Gamma$; this issue can be seen in the simpler case of functions in the presence of phase variation, wherein exact recovery is possible only when the dimensionality of phase variation is rendered finite-dimensional \citep{chakraborty2021functional, kurtek2011signal}. It may thus be possible to prove consistency by assuming a finite-dimensional parametric family of reparameterizations, but we leave that for future work. 

\cite{ghorbani2021functional} considered marked point processes with marks in an infinite-dimensional linear vector space in the absence of any symmetries and invariances, and proved consistency of an estimator of a $K$ function without the normalizing factor $c_\ff$. They assumed ergodicity of $\Psi$ and a known ground intensity function. However, their proof techniques cannot be used in our setting involving symmetries.

\section{Additional annotated tissue images}

\begin{figure}[!t]
    \centering
    \begin{tabular}{@{}c@{}}
         \includegraphics[width=0.8\linewidth]{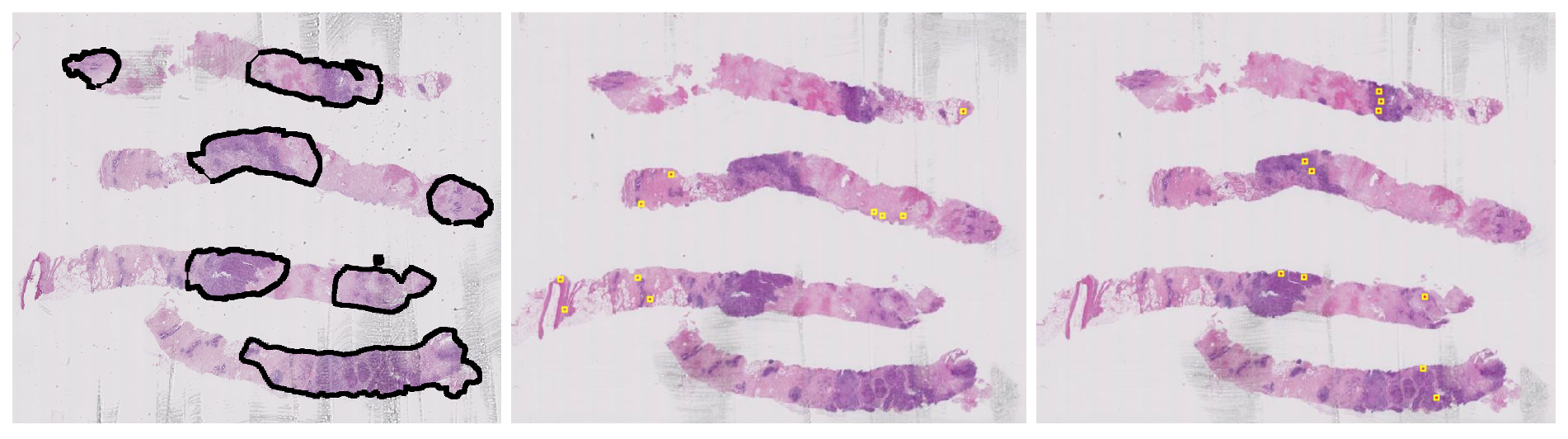} \\
         \includegraphics[width=0.8\linewidth]{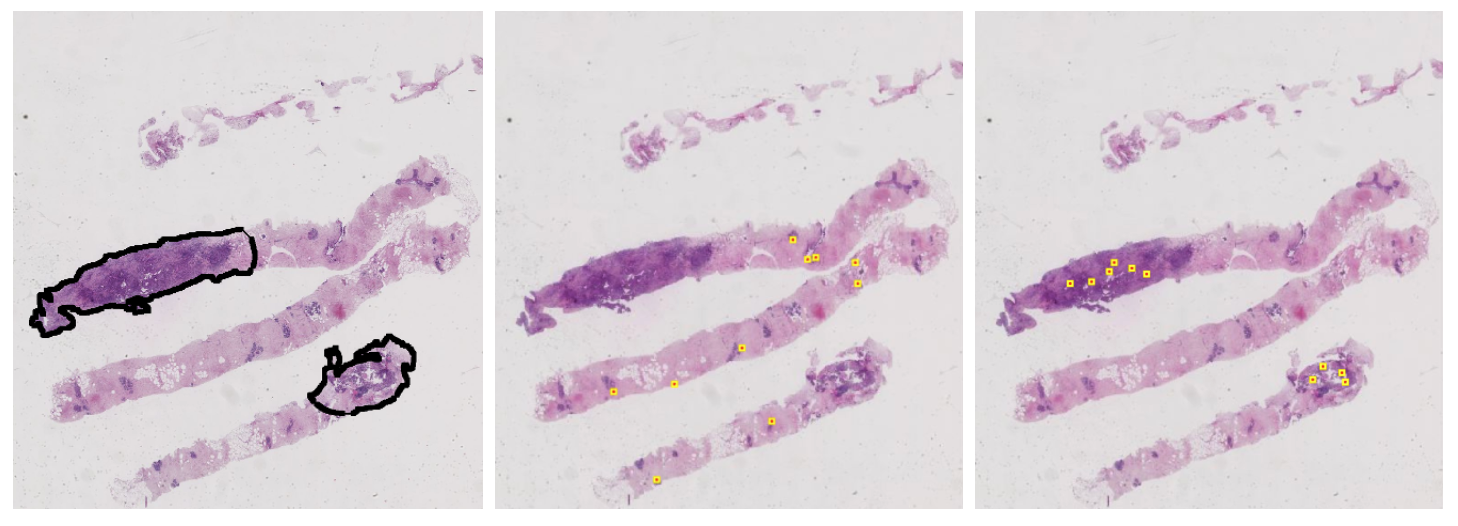} \\\includegraphics[width=0.8\linewidth]{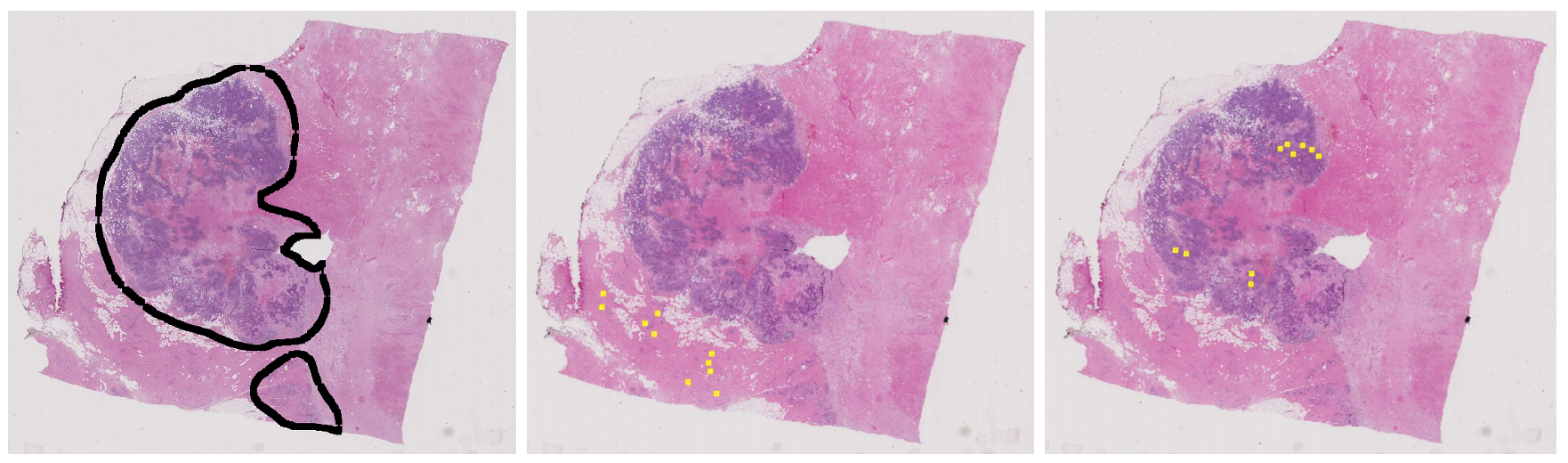} \\\includegraphics[width=0.8\linewidth]{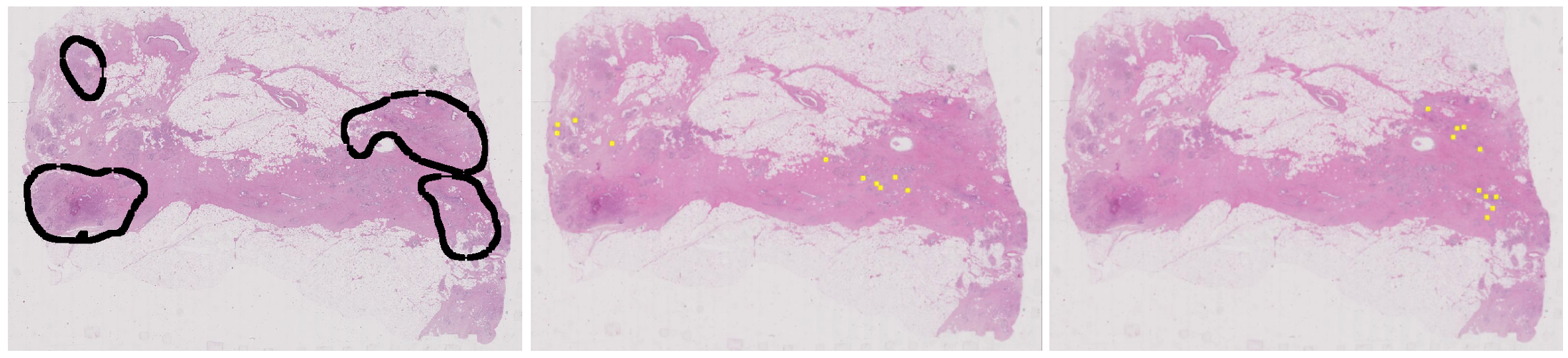} \\
    \end{tabular}
   \caption{\small Left: Annotated tissue image: tumor area is marked by the black boundary. Ten regions (yellow squares) were randomly sampled within the non-tumor area (middle) and tumor area (right).}
    \label{fig:breast-cancer-tissue-others}
\end{figure}

Figure \ref{fig:breast-cancer-tissue-others} displays the other four histopathology images that were used for the analysis in Section 7 in the main article. 

\section{Result for a region within non-tumor tissue}

\begin{figure}[!t]
    \centering
    \begin{tabular}{@{}c@{}c@{}c@{}}
         \multicolumn{3}{c}{\includegraphics[width=0.3\linewidth]{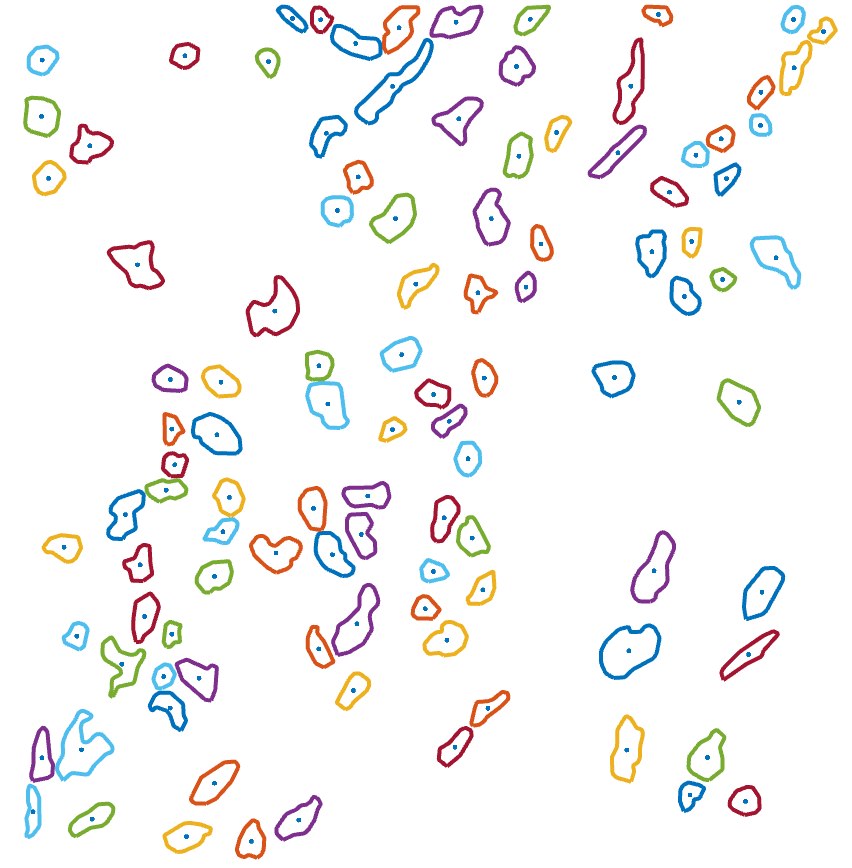 }} \\ 
         \multicolumn{3}{c}{(a)}\\
         \includegraphics[width=0.33\linewidth]{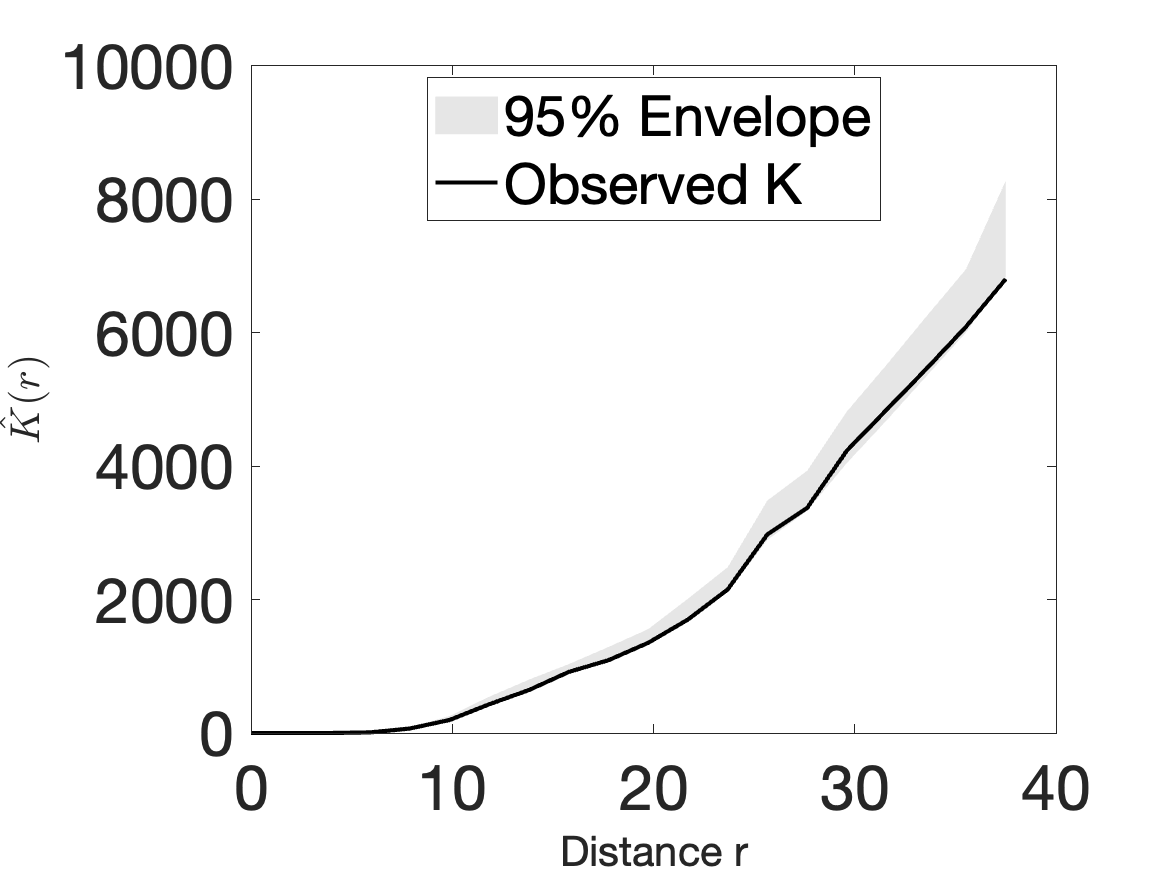} &
         \includegraphics[width=0.33\linewidth]{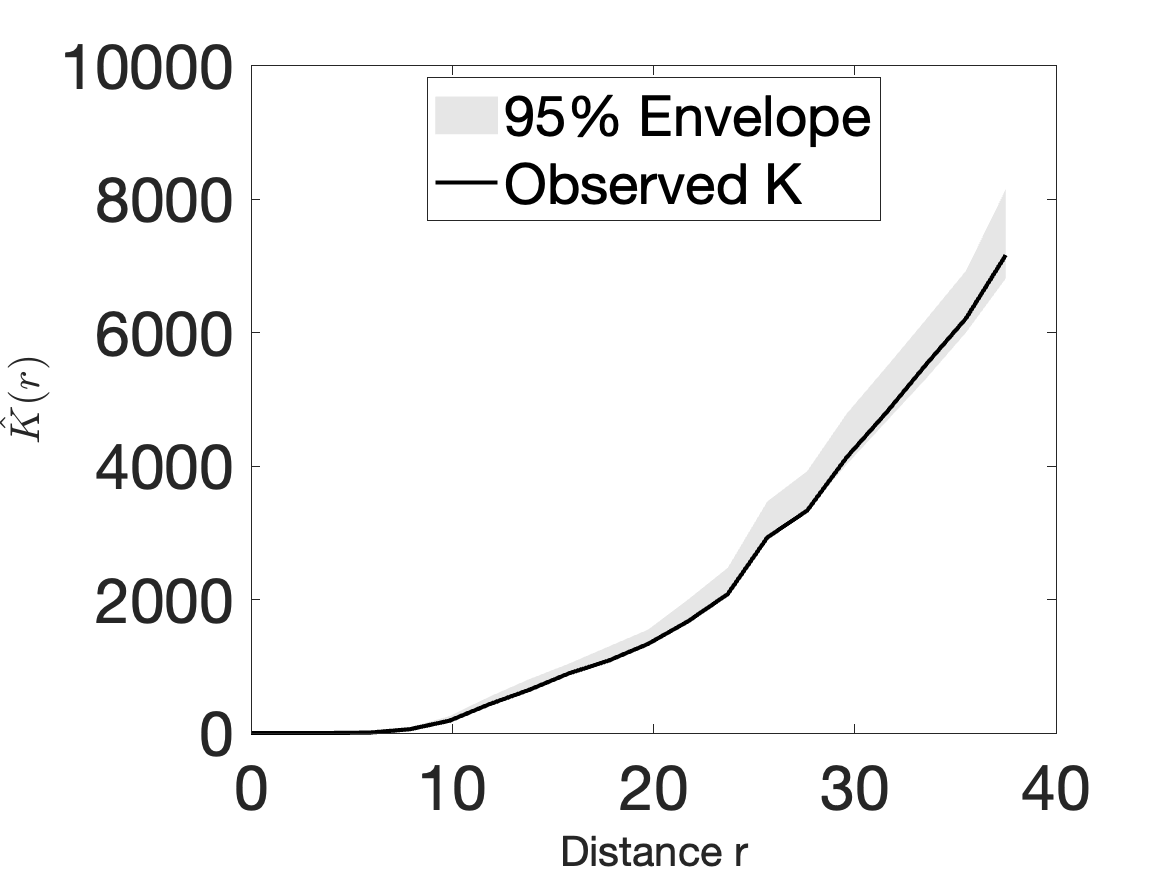} &
         \includegraphics[width=0.33\linewidth]{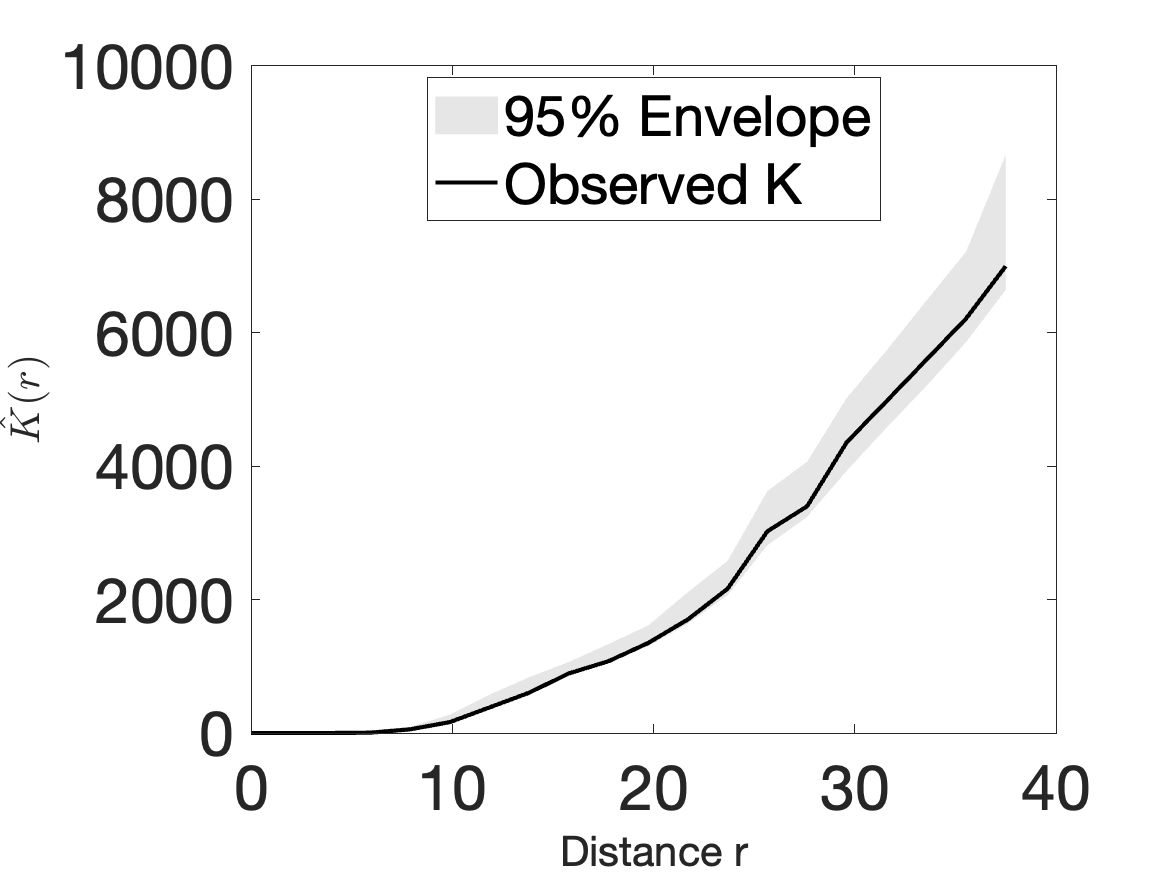}\\
         (b)&(c)&(d)\\
         \includegraphics[width=0.33\linewidth]{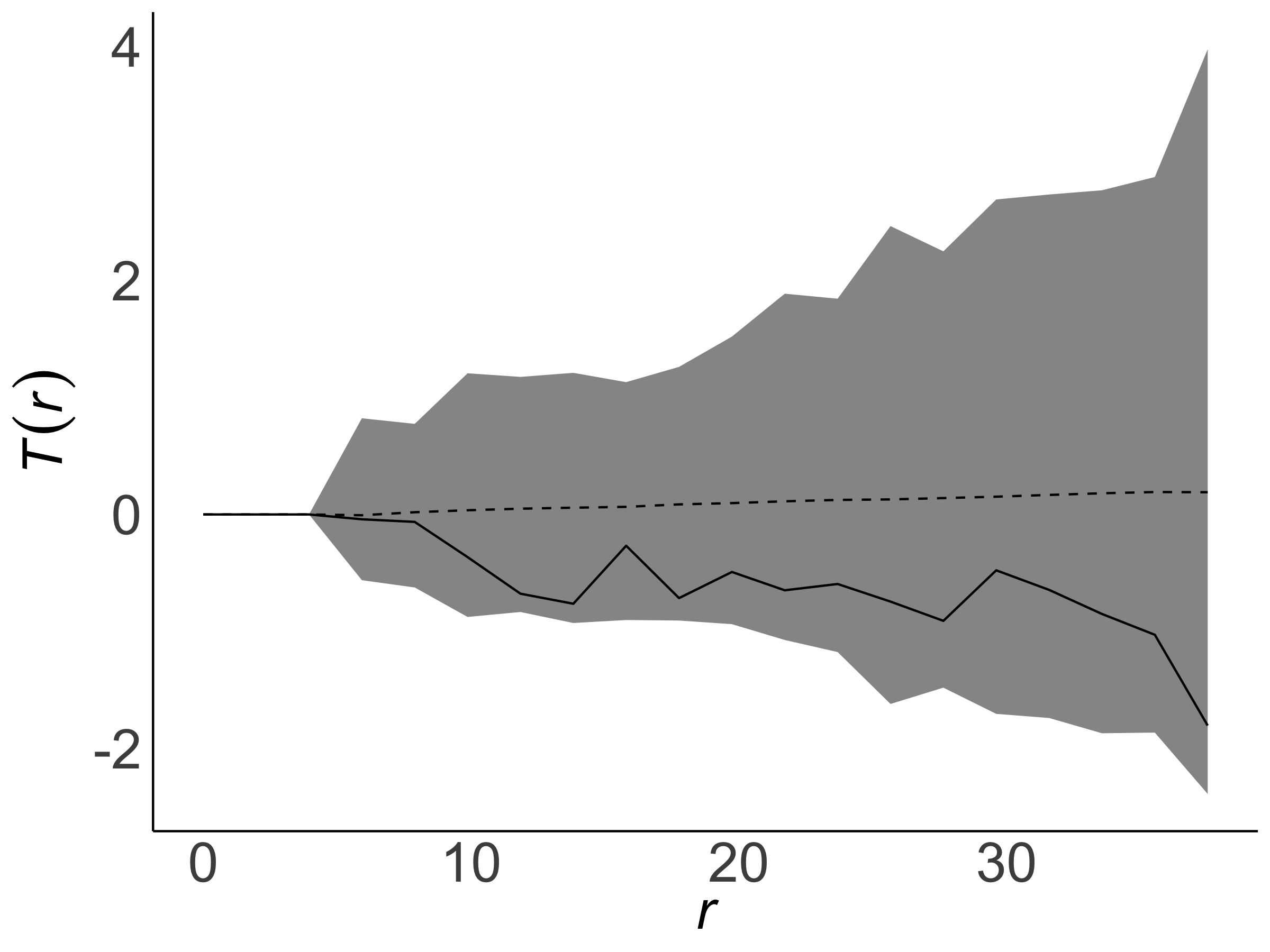} & \includegraphics[width=0.33\linewidth]{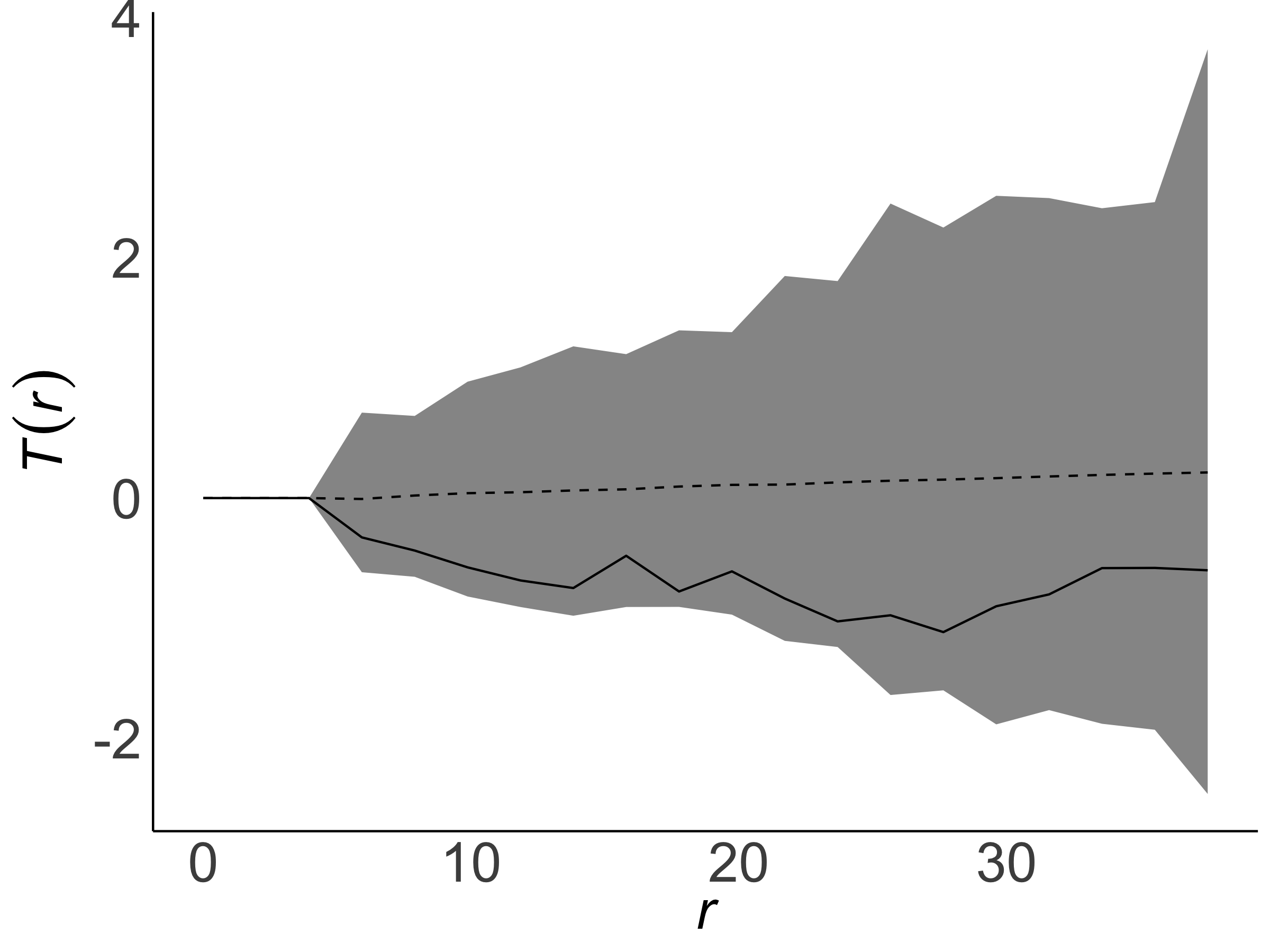} &
         \includegraphics[width=0.33\linewidth]{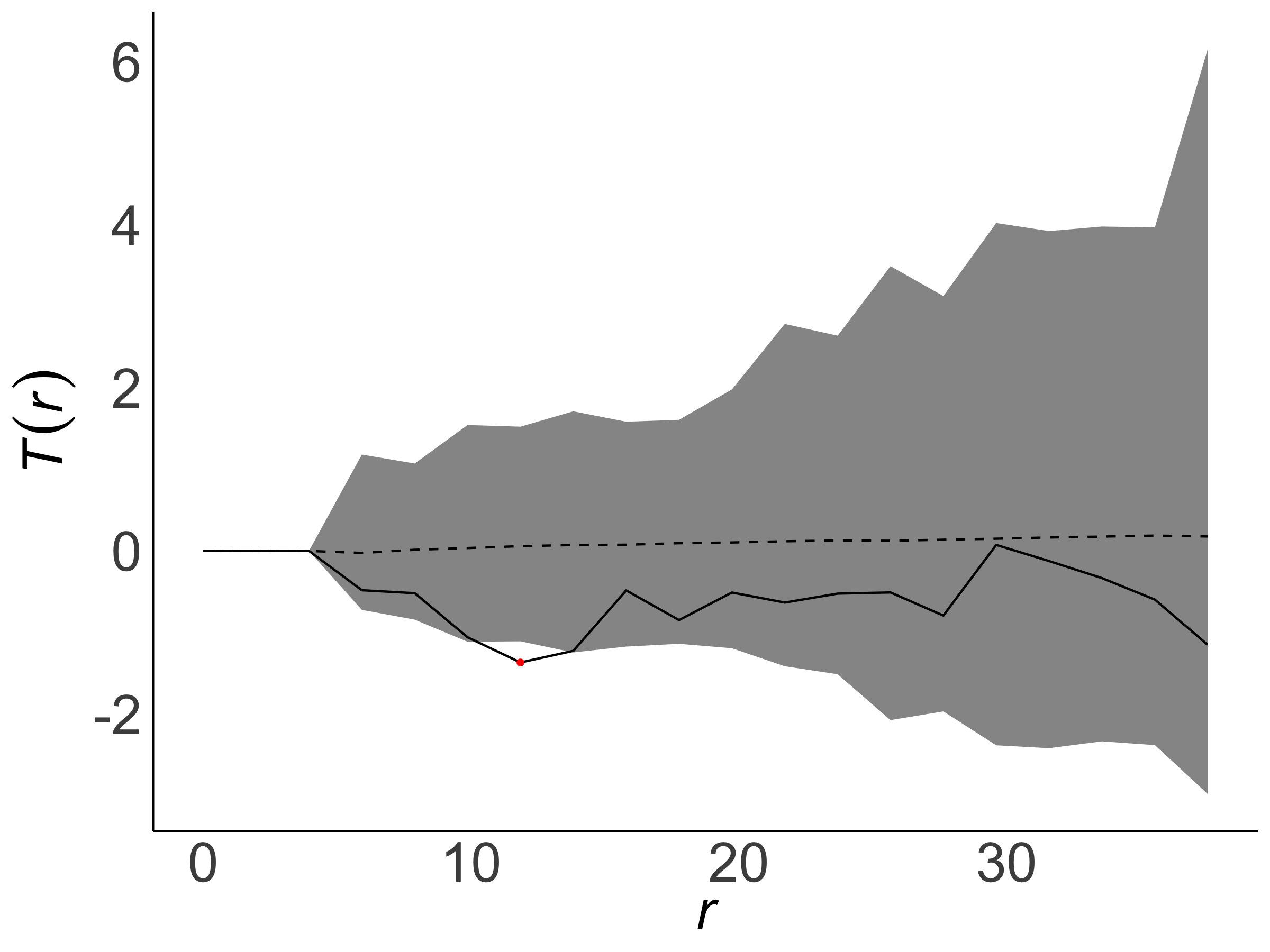}\\
         (e)&(f)&(g)
    \end{tabular}
   \caption{\small (a) A region within non-tumor tissue. Mark-weighted $K$ and $L$ functions along with their pointwise
and global envelopes, respectively, estimated using (b)\&(e) shape, (c)\&(f) orientation-and-shape,
and (d)\&(g) size-and-shape test functions.}
    \label{fig:k-breast-cancer-normal}
\end{figure}

Figure \ref{fig:k-breast-cancer-normal} presents results akin to Figure 6 in the main manuscript, but for a region sampled in non-tumor tissue. Here, we see that there is no strong evidence against random labeling based on shape, orientation-and-shape or size-and-shape. This observation was consistent for most regions sampled within non-tumor tissues, and supports the clinical interpretation of our results presented in Section 7 in the main article.

\bibliographystyle{plainnat}
\bibliography{ref}